\documentclass{aa}
\usepackage[varg]{txfonts}
\usepackage{graphicx}
\usepackage{hyperref}
\usepackage{multirow}
\usepackage{booktabs}
\usepackage{mathabx}
\usepackage{mathrsfs}

\begin{document}

\title{The backbone-residual model}
\subtitle{Accurately characterising the instrumental profile of a fibre-fed echelle spectrograph}

\author{Zhibo Hao\inst{1, 2, 4} \thanks{e-mail: zbhao@niaot.ac.cn}
\and Huiqi Ye \inst{1, 2}
\and Jian Han \inst{1, 2}
\and Liang Tang \inst{1, 2}
\and Yang Zhai \inst{1, 2}
\and Dong Xiao \inst{1, 2} \thanks{e-mail: dxiao@niaot.ac.cn}
\and Yongtian Zhu \inst{1, 2, 4}
\and Kai Zhang \inst{1, 2}
\and Liang Wang \inst{1, 2}
\and Gang Zhao \inst{3}
\and Fei Zhao \inst{3}
\and Huijuan Wang \inst{3, 4}
\and Jie Zheng \inst{3}
\and Yujuan Liu \inst{3}
\and Jiaqi Wang \inst{3, 4}
\and Ruyi Wei \inst{5}
\and Qiangqiang Yan \inst{5}}

\institute{
National Astronomical Observatories / Nanjing Institute of Astronomical Optics \& Technology, Chinese Academy of Sciences, Nanjing 210042, China
\and CAS Key Laboratory of Astronomical Optics \& Technology, Nanjing Institute of Astronomical Optics \& Technology, Nanjing 210042, China
\and CAS Key Laboratory of Optical Astronomy, National Astronomical Observatories, Beijing 100101, China
\and University of Chinese Academy of Sciences, Beijing 100049, China
\and CAS Key Laboratory of Spectral Imaging Technology, Xi’an institute of Optics and Precision Mechanics, Xi’an 710119, China
}

\date{}
\abstract
{
Instrumental profile (IP) is the basic property of a spectrograph.
Accurate IP characterisation is the prerequisite of accurate wavelength solution.
It also facilitates new spectral acquisition methods such as the forward modeling and deconvolution. 
}
{
We investigate an IP modeling method for the fibre-fed echelle spectrograph with the emission lines of the ThAr lamp, and explore the method to evaluate the accuracy of IP characterisation.
}
{
The backbone-residual (BR) model is put forward and tested on the fibre-fed High Resolution Spectrograph (HRS) at the Chinese Xinglong 2.16-m Telescope, which is the sum of the backbone function and the residual function.
The backbone function is a bell-shaped function to describe the main component and the spatial variation of IP. 
The residual function, which is expressed as the cubic spline function, accounts for the difference between the bell-shaped function and the actual IP. 
The method of evaluating the accuracy of IP characterisation is based on the spectral reconstruction and Monte Carlo simulation.
}
{
The IP of HRS is characterised with the BR model, and the accuracy of the characterised IP reaches 0.006 of the peak value of the backbone function.
This result demonstrates that the accurate IP characterisation has been achieved on HRS with the BR model, and the BR model is an excellent choice for accurate IP characterisation of fibre-fed echelle spectrographs.
}
{}
\maketitle

\section{Introduction}

An echelle spectrograph is able to realise a high-resolution ($> 10^{4}$) spectral observation of the celestial objects.
It plays a key role in many advanced astronomical researches, such as the detection of exoplanets with the radial velocity method and the measurement of isotopic ratios of stars.
The instrumental profile (IP) of an echelle spectrograph is the monochromatic response function in the one-dimensional extracted spectrum. 
Accurate IP characterisation supports a variety of applications, including but not limited to: 
\begin{enumerate}
\item the inspection of optical aberrations; 
\item the accurate determination of the absolute centres of spectral lines; 
\item the fitting of overlapping spectral lines; 
\item the spectral forward modeling; 
\item the spectral extraction with the method of deconvolution.
\end{enumerate}
Most importantly, it will facilitate accurate measurement of the celestial objects' spectra and spectral features (e.g. line equivalent width, line depth).
If the echelle spectrograph is fibre-fed, the fibre scrambler guarantees a rather stable IP for varied illumination on the fibre input face.
Therefore, the IP characterisation is not necessary to be simultaneous with the observation.
We can preform independent IP characterisation with calibration light sources for direct IP sampling.
The thorium argon (ThAr) lamp, which is a widely-used wavelength calibration source offering a forest of narrow emission lines in the visible and near-infrared range, can be employed for independent IP characterisation.

For realising an accurate IP characterisation, the method to model the IP is critical.
The simplest IP modeling method is to use the Gaussian function to fit the spectral lines.
Some bell-shaped functions (e.g. the super Gaussian function, the Moffat function and the Voigt function) are also commonly engaged.
\cite{martin2005} and \cite{zhao2014} suggested to fit the spectral lines with the sum of two Gaussian functions and achieved further improvement.
In the scenario of iodine technique, the widely-used California Planet Survey (CPS) Doppler code makes use of the sum of 12 Gaussians to model the IP \citep{valenti1995}.
CPS Doppler code has an alternative method of decomposing the IP by Gauss-Hermite bases \citep{wang2016}.
\cite{wang2016} proposed using a modified Moffat function $[1 + (x/\theta)^{2}]^{-\beta(\frac{x}{\delta})^{2}}$ to characterise the IP.
Due to similarity to IP modeling, the point spread function (PSF) modeling methods in the area of multi-fibre spectrographs are also noteworthy.
\cite{bolton2010} used the sum of a Gaussian function and a wing function to model the PSF.
\cite{kos2018} and \cite{bh2017} performed the PSF expansion by the basis functions of the discrete Chebyshev polynomial. 
\cite{li2015} fitted the PSF with a uniform B-spline surface.
Two-dimensional Gauss-Hermite bases were used in the work of \cite{cornachione2019}.
All the above modeling methods are based on a functional form or a combination of basis functions.
While A functional form can quickly and effectively capture the main component of IP, it is not universally accurate for all conditions.
A combination of basis functions is flexible and is able to capture the details of IP, but the components of high-order basis functions usually lead to unwanted high-frequency oscillation in the characterised IP.
Even though the latter can be avoided with smooth piecewise functions such as the cubic spline function, the need of combining a large amount of data to outline the curve of IP with a high sampling rate poses challenges for practical applications.
For example, an iterative process of repeatedly adjusting the parameters for centreing \citep{anderson2000} would be necessary.

The purpose of this work is to develop a generalised accurate IP modeling method for fibre-fed echelle spectrographs with the emission lines of the ThAr lamp.
This modeling method can achieve accurate IP characterisation, while being free of cumbersome processes like iteration.
We demonstrates that a model combining a bell-shaped function and a cubic spline function can meet our goal.
And an IP characterisation test based on our method is carried out on the fibre-fed High Resolution Spectrograph (HRS) of the Chinese Xinglong 2.16-m Telescope.
In addition, by using the spectra of the astro-comb equipped on HRS, we evaluate the accuracy of IP characterisation. 
The rest of this paper is structured as follows: 
The preliminary analysis about the IP characterisation is given in Section 2.
In Section 3 we introduce our IP modeling method and present the process of IP characterisation on HRS with the exposures of ThAr lamp, and the result is shown in Section 4.
In Section 5, we explore the method of evaluating the accuracy of IP characterisation, and apply it to our IP result.
Brief discussion and conclusions are given in Section 6.

\section{Preliminary analysis}

\subsection{Anamorphic tangential magnification}

For an echelle spectrograph, the spatial variation of IP is mostly caused by the anamorphic tangential magnification \citep{christopher2005}, which is produced by the dispersion of the echelle grating. 
This is particularly true when most of the optical aberrations are well suppressed by the well-designed camera lens system.
We perform a simple theoretical analysis below to explain this point and estimate the level of its influence.
 
\begin{figure}
\centering
\resizebox{\hsize}{!}{\includegraphics{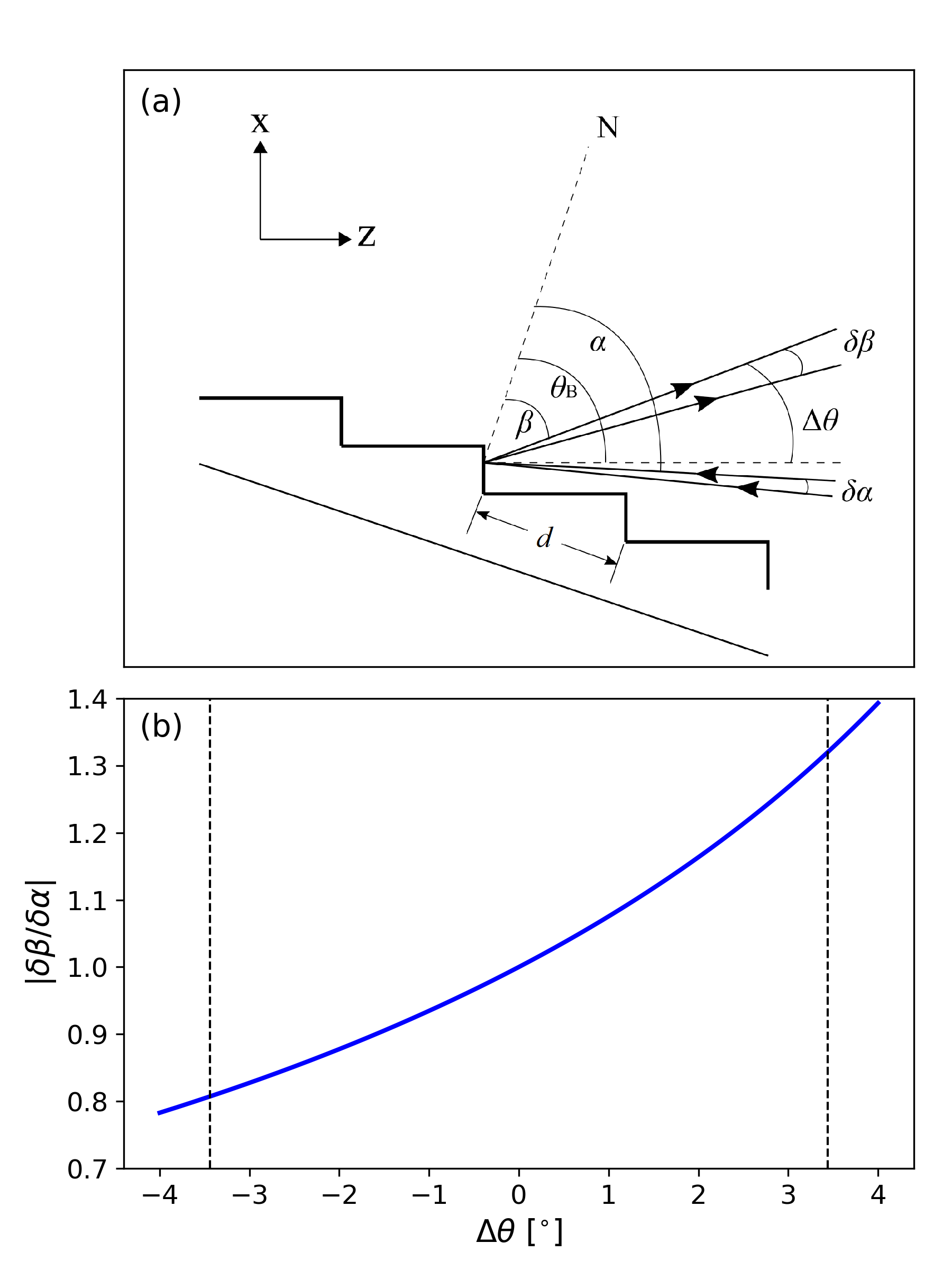}}
\caption{
{\bfseries a)} Geometry showing the anamorphic tangential magnification. 
Monochromatic light from the fibre output face or the entrance slit, of the subtended angle $\delta \alpha$, is projected by the echelle grating to form an image of the subtended angle $\delta \beta$.
The staircase-like solid thick lines indicate the cross-section of the echelle grating. 
See text for detailed descriptions of other symbols. 
{\bfseries b)} The absolute value of the ratio of $\delta \alpha$ to $\delta \beta$ as the function of the off-axis angle, given that the R4 echelle grating of the blaze angle $\theta_{B} = 76^{\circ}$ is used. 
The two dashed lines mark the positions of $\pm 3.44^{\circ}$, which indicate the off-axis angles of the edges of the CCD of the Ultraviolet-Visible Echelle Spectrograph (UVES).
}
\label{fig:2.1}
\end{figure}

Fig.\ref{fig:2.1}(a) shows a cross-section of an echelle grating with blaze angle $\theta_{B}$. 
$\alpha$ and $\beta$ are the angles of incidence and diffraction, respectively.
All the rays are in the xz-plane.
The x-axis is along the principal dispersion direction, and the z-axis is normal to the reflection surface of the grooves.  
The diffraction relation of the echelle grating indicates \citep{schroeder1970, schroeder1987}: 
\begin{equation}
\frac{o\lambda}{d} = \rm{sin}\alpha + \rm{sin}\beta,
\label{eq:2.1}
\end{equation}
where $o$ is the order number of wavelength $\lambda$, and $d$ is the grating constant.
Assuming the angle subtended by the fibre output face or the entrance slit to the incident point on the groove surface is $\delta\alpha$, we obtain:
\begin{equation}
\frac{o\lambda}{d} = \rm{sin}(\alpha+\delta\alpha) + \rm{sin}(\beta+\delta\beta),
\label{eq:2.2}
\end{equation}
where $\vert\delta\beta\vert$ is the angle subtended by the corresponding image on the CCD. 
Therefore, the magnification of subtended angles satisfies:
\begin{equation}
\vert\frac{\delta\beta}{\delta\alpha}\vert = \vert\frac{\rm{cos}\alpha}{\rm{cos}\beta}\vert.
\label{eq:2.3}
\end{equation}
We assume the echelle spectrograph is operated in quasi-Littrow condition.
Hence, $\alpha \approx \theta_{B}$ and $\beta \approx \theta_{B}+\Delta\theta$.
Here, $\Delta\theta$ denotes the off-axis angle, i.e. the angle between the diffractive ray and the optical axis of the optical system.
Finally we obtain:
\begin{equation}
\vert\frac{\delta\beta}{\delta\alpha}\vert \approx \vert\frac{\rm{cos}\theta_{B}}{\rm{cos}(\theta_{B}+\Delta\theta)}\vert.
\label{eq:2.4}
\end{equation}
Eq. \ref{eq:2.4} presents the anamorphic tangential magnification: The magnification in the principal dispersion direction is related to the off-axis angle.
Assuming the R4 echelle grating is used, substituting its typical blaze angle $\theta_{B} = 76^{\circ}$ into Eq. \ref{eq:2.4}, the magnification as the function of $\Delta\theta$ is derived as shown in Fig.\ref{fig:2.1}(b) with the blue curve.
Different off-axis angles correspond to different positions on the CCD.
Taking the red camera of the Ultraviolet-Visible Echelle Spectrograph (UVES) as an example, the magnifications are 0.807 and 1.320 respectively for the right edge (off-axis angle = $-3.44^{\circ}$) and left edge (off-axis angle = $3.44^{\circ}$) of the CCD, given that the focal length is 500 mm and the CCD is 4K $\times$ 4K with a pixel size of 15 $\mu$m \citep{dekker2000}.
It means if we do not consider the contribution of optical aberrations, the width of IP at the left edge is $1.320/0.807=1.636$ times larger than that at the right edge. 
Such a difference is commonly much bigger than that induced by the optical aberrations of the echelle spectrograph.
Note that the magnification varies gradually along the principal dispersion direction over the whole range of CCD, rather than abruptly in the region of large off-axis angle.

\subsection{A selection criterion for the ThAr lines}

\begin{figure}
\centering
\resizebox{\hsize}{!}{\includegraphics{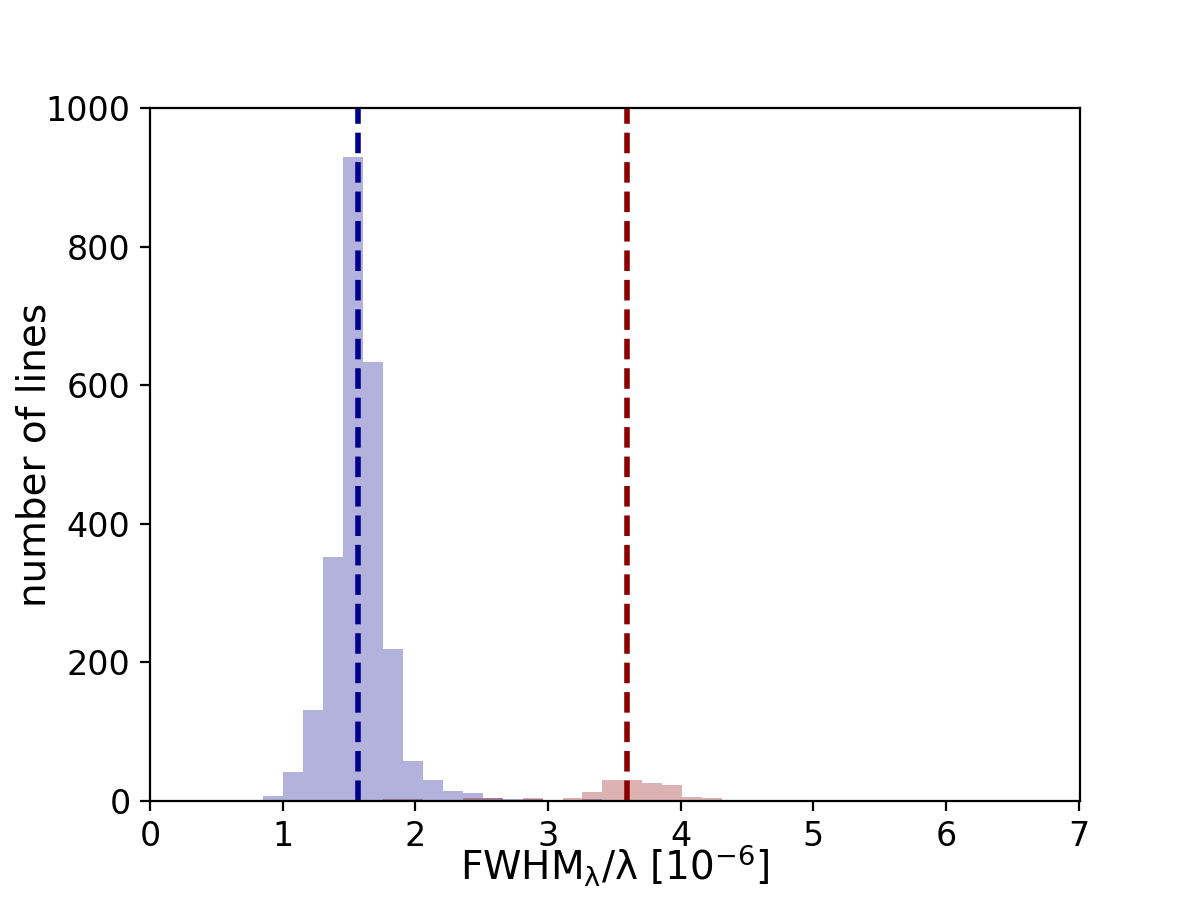}}
\caption{
Histogram of the relative intrinsic widths of the thorium lines (blue) and argon lines (red) of a ThAr lamp.
The data are taken from Table 1 of \cite{redman2014}.
The dashed lines indicate the medians of the relative intrinsic widths of the thorium lines (blue) and the argon lines (red).
}
\label{fig:2.2}
\end{figure}

Generally, an accurate IP characterisation is only achievable with spectral lines of sufficiently narrow intrinsic widths.
We take the data from Table 1 of \cite{redman2014}, which recorded the measurement results of ThAr lines with the National Institute of Standards and Technology (NIST) 2 m Fourier transform spectrometer in 2013, and investigate the distribution of intrinsic line widths. 
Fig.\ref{fig:2.2} shows the histogram of $\textrm{FWHM}_{\lambda}/\lambda$ for thorium (blue) and argon (red) emission lines, where $\textrm{FWHM}_{\lambda}$ denotes the intrinsic line widths and $\lambda$ denotes the wavelength. 
The blue dashed line indicates the median of $\textrm{FWHM}_{\lambda}/\lambda$ for the thorium lines, which is equal to $1.57 \times 10^{-6}$. 
The red dashed line indicates the median of $\textrm{FWHM}_{\lambda}/\lambda$ for the argon lines, which is equal to $3.59 \times 10^{-6}$. 

The measured profile of a spectral line is produced by convolving its intrinsic line profile with the IP. 
So, the width of the measured spectral line roughly satisfies the relation of quadratic sum: 
\begin{equation}
\textrm{FWHM}_{\lambda, \rm{measured}}^{2} \sim (\lambda/R)^{2} + \textrm{FWHM}_{\lambda}^{2},
\label{eq:5}
\end{equation}
where $R$ is the resolving power of a spectrograph (so $\lambda/R$ is the width of IP).
We find $\textrm{FWHM}_{\lambda, \rm{measured}} < \sqrt{(\lambda/R)^{2} + (0.1\lambda/R)^{2}} \approx 1.005\lambda/R$ when the intrinsic line width is less than one tenth of the IP width. 
That means the difference between the measured line profile and the IP is less than 0.5\%. 
This level of error is acceptable.
We take it as a criterion in the selection of applicable emission lines for IP characterisation.
Based on the values of the median of $\textrm{FWHM}_{\lambda}/\lambda$, we find that more than half of the thorium lines or argon lines are applicable to IP characterisation when the resolving power of the spectrograph $R \leq 6.37 \times 10^{4}$ or $R \leq 2.79 \times 10^{4}$, respectively.
This result demonstrates the capability of ThAr lamp as a light source for direct IP characterisation for echelle spectrographs.
However, for very high-resolution spectrographs, e.g. HARPS ($R = 1.15 \times 10^{5}$) \citep{mayor2003}, the number of applicable ThAr lines would be quite limited.

\section{Characterising the IP}

\subsection{Backbone-residual (BR) model}

According to the aforementioned analysis, the spatial variation of IP of an echelle spectrograph is largely influenced by the anamorphic tangential magnification.
This kind of variation manifests itself as the broadening or compression of the shape of IP in the principal dispersion direction.
By fitting the IP with a bell-shaped function, broadening or compression can be well captured.
In this way, the spatial variation of IP remaining in the residuals (i.e. the difference between the best-fit bell-shaped function and the actual IP) is largely reduced.
These residuals, which reveal the details of IP, thus have better internal consistency of appearing as the same function than the whole IP. 
The systematic errors included in the combined residuals will be smaller than those included in the directly-combined IP.
Moreover, using the best-fit bell-shaped function to fit the applicable ThAr lines, we can obtain the peak heights and centres of these lines.
Taking them as the normalisation constants and the centres of IP, the processes of normalisation and centreing before data combination can be easily implemented, free of the cumbersome processes like iteration.
Therefore, it is beneficial to include the best-fit bell-shaped function in the model of IP characterisation.
On the other hand, the cubic spline function is very flexible and is always able to accurately trace an arbitrary smooth function without high-frequency oscillation.
Therefore, the residuals, which commonly have an undulating structure, can be well described by the cubic spline function.
In short, we propose the backbone-residual (BR) model:
\begin{equation}
\textrm{IP}(x'; o, x) = \textrm{IP}_{\textrm{b}}(x'; o, x) + \textrm{IP}_{\textrm{r}}(x'; o, x).
\label{eq:6}
\end{equation}
$x'$ is the argument of the model. 
It represents the pixel position relative to the centre of IP.
The order $o$ and the pixel position $x$ in the principal dispersion direction are the parameters which present the two-dimensional position of IP on the CCD.
The BR model is the sum of the backbone function $\textrm{IP}_{\textrm{b}}(x'; o, x)$ and the residual function $\textrm{IP}_{\textrm{r}}(x'; o, x)$.
The backbone function is the best-fit bell-shaped function, which is named after its role of accounting for the main component of IP. 
The residual function, which is expressed as the cubic spline function, describes the difference between the backbone function and the actual IP.
For normalisation, we define the peak value of the backbone function as equal to 1.
So, the residual function represents the relative values of residuals as to the peak value of the backbone function.
We would like to address that our definition of normalisation is unconventional: the integral of IP is NOT equal to 1.
Attention should be paid to the effects caused by the definition of normalisation when applying the BR model to the operations such as the spectral reconstruction.

\subsection{Test-system description}

We test IP characterisation with the BR model on HRS.
HRS \citep{fan2016} is a fibre-fed echelle spectrograph with a wavelength coverage of 370-920 nm. 
The detector is a 16-bit back-illuminated E2V CCD (203–82), with a size of 4096 $\times$ 4096 pixels and a pixel-size of 12 $\mu$m.
The average resolving power is $4.3 \times 10^{4}$ for the 2.4'' multi-mode fibre which transmits the the light of target from the Cassegrain unit to the spectrograph.
The average width of IP is 5.5 pixels.
An iodine absorption cell and a ThAr lamp are equipped as wavelength calibrators.
A new generation wavelength calibration source -- an astro-comb (Menlo Systems GmbH) was installed on HRS in 2016 \citep{ye2016}. 

The IP characterisation utilises the exposures of ThAr lamp on HRS.
Prior to the IP characterisation, basic data reduction needs to be carried out, including bias subtraction, scattered light subtraction, order tracing, order extraction, wavelength calibration, etc.
And finally, the one-dimensional spectrum and the position of each ThAr line are obtained.
The data reduction pipeline is developed according to the general techniques for echelle spectrographs (e.g. \cite{buchhave2010}).
We note that we extract the one-dimensional spectrum by directly summing up $\pm7$ pixels in the cross-dispersion direction.

\subsection{The determination of the backbone function and the residual function for HRS}

The determination of the backbone function should be based on the actual shape of the IP of spectrograph. 
The test on HRS suggests the normalised super Gaussian function as the backbone function: 
\begin{equation}
\textrm{IP}_{\textrm{b}}(x'; o, x) = \exp[-(\frac{\vert x' \vert}{\sigma(o, x)})^{\beta(o, x)}],
\label{eq:12}
\end{equation}
because there are only two parameters $\sigma$ and $\beta$ controlling the shape while the residuals are less than 5\% of the height everywhere.
This performance is better than other common bell-shaped functions (e.g. the Gaussian function, the Moffat function and the Voigt function) on HRS.

The residual function is expressed as the cubic spline function:
\begin{equation}
\textrm{IP}_{\textrm{r}}(x'; o, x) = \sum_{j=1}^{n} c_{j}(o, x)B_{j, d}(x'),
\label{eq:13}
\end{equation}
where $n$ is the number of coefficients and $d=3$.
The B-spline basis function $B_{j, d}(x')$ is defined as the iterative form:
\begin{equation}
B_{j, d}(x') = \frac{x'-x'_{j}}{x'_{j+d}-x'_{j}}B_{j, d-1}(x') + \frac{x'_{j+d+1}-x'}{x'_{j+d+1}-x'_{j}}B_{j+1, d-1}(x'), 
\label{eq:14}
\end{equation}
\begin{equation}
B_{j, 0}(x') = \left\{ 
               \begin{array}{ll}
               1, & x'_{j} \leq x' < x'_{j+1} \\
               0, & x' < x'_{j} \ \textrm{or} \ x' \geq x'_{j+1} \\
               0, & \textrm{if} \ x'_{j} = x'_{j+1}. \\
               \end{array}
\right.
\label{eq:15}
\end{equation}
$ \{ x'_{j} \}_{j=1}^{n+d+1}$ compose the set of knots.

Consequently, the parameters in the BR model for HRS consist of $\sigma$, $\beta$ of the super Gaussian function and $\{ c_{j} \}_{j=1}^{n}$ of the cubic spline function. 
The IP characterisation is thus essentially to obtain the parameters as the functions of order $o$ and pixel position $x$ by using the actual exposures of ThAr lamp.
Fitting each applicable ThAr line with a super Gaussian function yields the best-fit $\sigma$ and $\beta$ at the position.
However, in general, the data points of the residuals of a single ThAr line only supply a poor sampling rate in the fitting range which can hardly reveal the undulating structure of the residual function.
This can be solved by combining the normalised and centred residuals of the ThAr lines of different pixel phases in a particular area (see Section 3.4), into a united set of data points to achieve a high sampling rate.

\subsection{The process of IP characterisation}

\begin{figure}
\centering
\resizebox{\hsize}{!}{\includegraphics{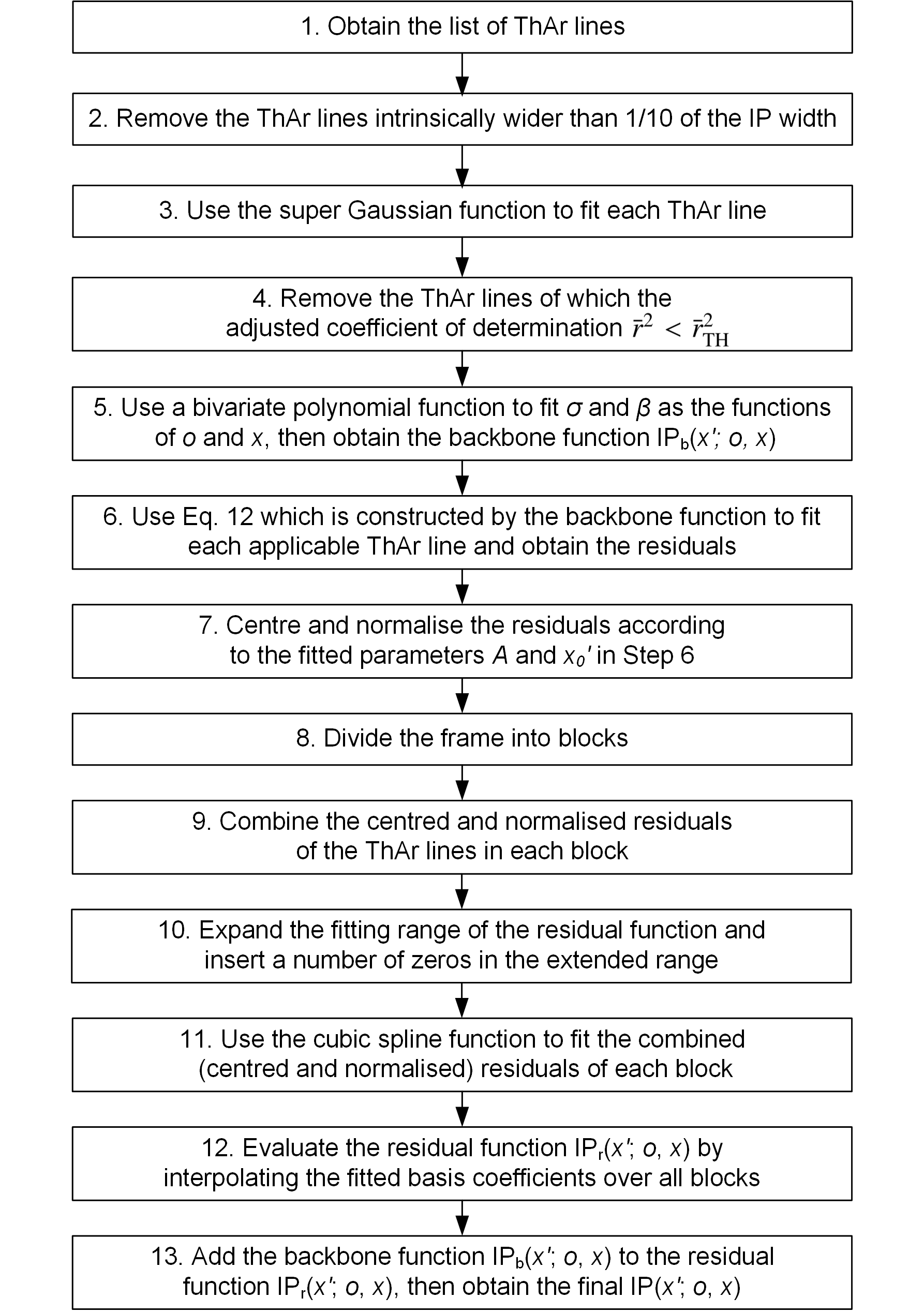}}
\caption{
Flowchart of the process of IP characterisation with an actual exposure of ThAr lamp on HRS.
}
\label{fig:3.1}
\end{figure}

\begin{figure*}
\centering
\resizebox{\hsize}{!}{\includegraphics{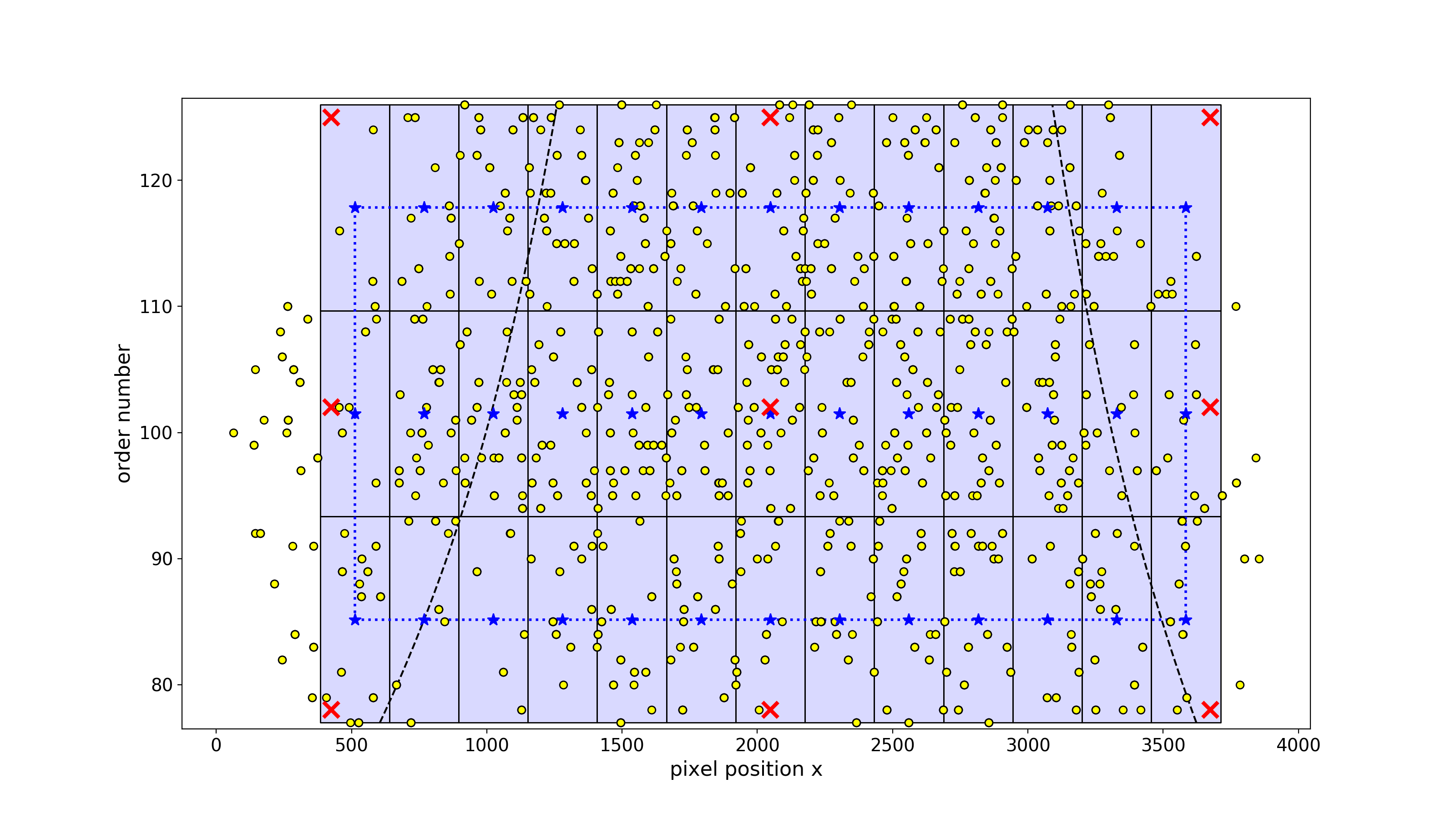}}
\caption{
This figure shows the block division (divided by the solid black lines) with respect to the combination of the (centred and normalised) residuals of ThAr lines in the case of HRS.
The combination is performed separately with the (centred and normalised) residuals in each block.
In this paper, we use the order $o$ to describe the coordinates in the cross-dispersion direction, instead of the pixel position. 
An example of the distribution of applicable ThAr lines of a ThAr exposure on HRS is shown by the yellow circles.
The blue stars mark the centres of blocks.
The convex hull of block centres is shown as the dotted blue lines.
The non-repetitive parts of orders (i,e. the parts having unique wavelength coverage) are shown as the area between the two dashed black lines.
See the caption of Fig.\ref{fig:4.1} for the description of the red crosses. 
}
\label{fig:3.2}
\end{figure*}

The diagram shown in Fig.\ref{fig:3.1} presents the flowchart of the process of IP characterisation with an actual exposure of ThAr lamp on HRS.
There are 13 steps in total.
Although a few operations and settings are specific to HRS, this process can be generally broadened to other fibre-fed echelle spectrographs with only slight modifications.

The goal of the first five steps is to obtain the parameters ($\sigma$ and $\beta$) of the backbone function as the functions of order $o$ and pixel position $x$, including two steps to clip out the ``bad'' ThAr lines.
Those ``bad'' ThAr lines either have too broad intrinsic widths to pass the width criterion defined in Section 2.2 (clipped out in Step 2), or have too large residuals from the fitted super Gaussian function (clipped out in Step 4).
In Step 4, the large residuals may be caused by the disturbance of the unseparated closely-neighboring spectral lines, or merely by the low SNR of lines.
We use the adjusted coefficient of determination $\bar{r}^{2}$ to test the goodness of fit.
It is defined as:
\begin{equation}
\bar{r}^{2} = 1 - \frac{\textit{SS}_{\textrm{res}}/(n-p-1)}{\textit{SS}_{\textrm{tot}}/(n-1)},
\label{eq:3.4.1}
\end{equation}
where $\textit{SS}_{\textrm{res}}$ denotes the residual sum of squares, $\textit{SS}_{\textrm{tot}}$ denotes the total sum of squares, $n$ and $p$ denote the number of data points and that of the free parameters, respectively. 
$\bar{r}^{2}_{\textrm{THOLD}}$ is the threshold for clipping.
The ThAr lines of $\bar{r}^{2} < \bar{r}^{2}_{\textrm{THOLD}}$ will be clipped out.
$\bar{r}^{2}_{\textrm{THOLD}}$ must be carefully determined to balance the effectiveness of clipping and the tolerance of the difference of the actual IP and the fitted super Gaussian function.
In the case of HRS, $\bar{r}^{2}_{\textrm{THOLD}}$ is set to 0.995.
After Step 4, we obtain the best-fit $\sigma$ and $\beta$ of each applicable ThAr line.
Then, a low-order 2D polynomial fitting is performed to determine $\sigma$ and $\beta$ as the functions of $o$ and $x$, i.e. $\sigma(o, x)$ and $\beta(o, x)$.
In the case of HRS, 1st-order in $o$ and 3rd-order in $x$ are found to be adequate.

The goal of Step 6 to 12 is to obtain the coefficients ($\{ c_{j} \}_{j=1}^{n}$) of the residual function as the functions of order $o$ and pixel position $x$.
In Step 6, we use the backbone function to construct a model function:
\begin{equation}
f(x'; o, x) = A\,\textrm{IP}_{\textrm{b}}(x'-x'_{0}; o, x) + C
\label{eq:3.4.2}
\end{equation}
to refit each applicable ThAr line and acquire the residuals.
Here $A$, $x'_{0}$ and $C$ denote the parameters representing the peak height, line centre and background. 
This operation is equivalent to use a background-added super Gaussian function to refit each line while keeping $\sigma = \sigma(o, x)$ and $\beta = \beta(o, x)$.
The fitted $A$ and $x'_{0}$ of each line are employed in Step 7 as the normalisation constant and the centre of IP.

As we mentioned in Section 3.3, we need to combine the centred and normalised residuals of different ThAr lines before fitting the residual function.
In the case of HRS, the combination is performed separately for each block as shown in Fig.\ref{fig:3.2} (divided by the solid black lines).
The way of block division is optimised to balance between the high sampling rate (tending to enlarge the area of each block) and internal consistency of the combined residuals (tending to reduce the area of each block).
Section 2.1 shows the spatial variation caused by the anamorphic tangential magnification mainly occurs in the principal dispersion direction.
Therefore, the block division in the principle dispersion direction is dense (256-pixel wide) while there are only three rows in the cross-dispersion direction.
The (centred and normalised) residuals of the applicable ThAr lines within each block are combined to form a united dataset for the residual function fitting.
Fig.\ref{fig:3.2} also shows an example of the distribution of applicable ThAr lines by the yellow circles.
Due to the decline of the diffraction efficiency towards the border of the CCD, the applicable ThAr lines close to the edges are quite limited.
Therefore, owing to the lack of data, we omit regions close to the edges. 
In spite of this, the non-repetitive parts of orders (i,e. the parts having unique wavelength coverage, shown as the area between the two dashed black lines in Fig.\ref{fig:3.2}) are still entirely covered by the blocks.

The fitting range of IP is set as $x' \in [-7.5, 7.5]$ based on the test on HRS which shows the IP converges to zero inside [-7.5, 7.5] anywhere on the CCD.
Meanwhile, we note that it is beneficial to introduce buffer spaces for lessening the influence of boundary condition on the shape of the fitted spline function. 
Therefore, in Step 10, we expand the fitting range of the residual function to $x' \in [-10, 10]$.
And a number of zeros are artificially inserted in the intervals of [-10, -7.5] and [7.5, 10]. 

In Step 11, the cubic spline function fitting is carried out on the combined (centred and normalised) residuals of each block.
In the case of HRS, for constructing the spline function, we arrange 21 interior knots in the fitting range [-10, 10] (so the number of coefficients is $n=21+3+1=25$).
Note that the knots are not uniformly distributed in the fitting range.
They are arranged more densely in [-5, 5], where the residual function is undulating, and more sparsely outside [-5, 5] (see Appendix A for the details).
After iterative fitting and 3$\sigma$-clipping, the 25 coefficients of fitted spline function of each block are obtained.
They are regarded as the representation of the residual function based on B-spline basis functions at the {\bfseries centre} of each block (blue stars in Fig.\ref{fig:3.2}). 
Then, Step 12 implements linear interpolation on each coefficient inside the convex hull of block centres, and implements nearest-neighbor interpolation outside the convex hull of block centres.
 $\{ c_{j}(o, x) \}_{j=1}^{n}$ as the functions of order $o$ and pixel position $x$ are thus all obtained.
The convex hull of block centres is shown as the dotted blue lines in Fig.\ref{fig:3.2}.
Finally, according to Eq. \ref{eq:13}, $\{ c_{j}(o, x) \}_{j=1}^{n}$ and B-spline basis functions are joined to build the residual function $\textrm{IP}_{\textrm{r}}(x'; o, x)$. 

So far, both the backbone function and the residual function have been obtained. 
In Step 13, the sum of them gives rise to the final IP expression, i.e., $\textrm{IP}(x'; o, x)$. 

If more than one ThAr exposure is taken in a short time, the data points of different exposures can be combined when we implement the $\sigma(o, x)$ and $\beta(o, x)$ fitting and the residual function fitting.
Because in a short time, the change of IP caused by the instability of the illumination or the instrumental environment can be ignored.
Instead, the spectrograph drift produces the dithering of ThAr lines that provides more adequate IP sampling on the CCD.  
Taking advantage of the combined data, the accuracy of IP characterisation would be further improved.

\section{The result of IP characterisation}

\begin{figure*}
\centering
\resizebox{\hsize}{!}{\includegraphics{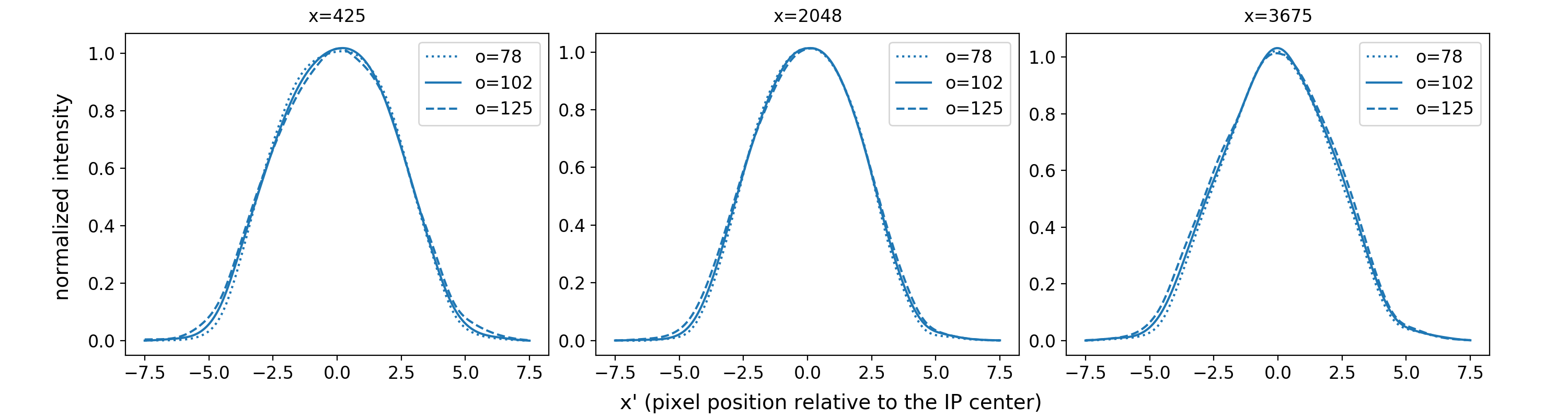}}
\caption{
The result of IP characterisation based on the BR model at nine positions on the CCD of HRS.
These nine positions are marked in Fig.\ref{fig:3.2} as the red crosses. 
They correspond to the four corners, the centres of four sides, and the centre of the region of blocks, respectively.
Each subfigure shows the IP of the same pixel position $x$ (marked on the top).
The IP of different orders are distinguished by the line style.
}
\label{fig:4.1}
\end{figure*}

\begin{figure}
\centering
\resizebox{\hsize}{!}{\includegraphics{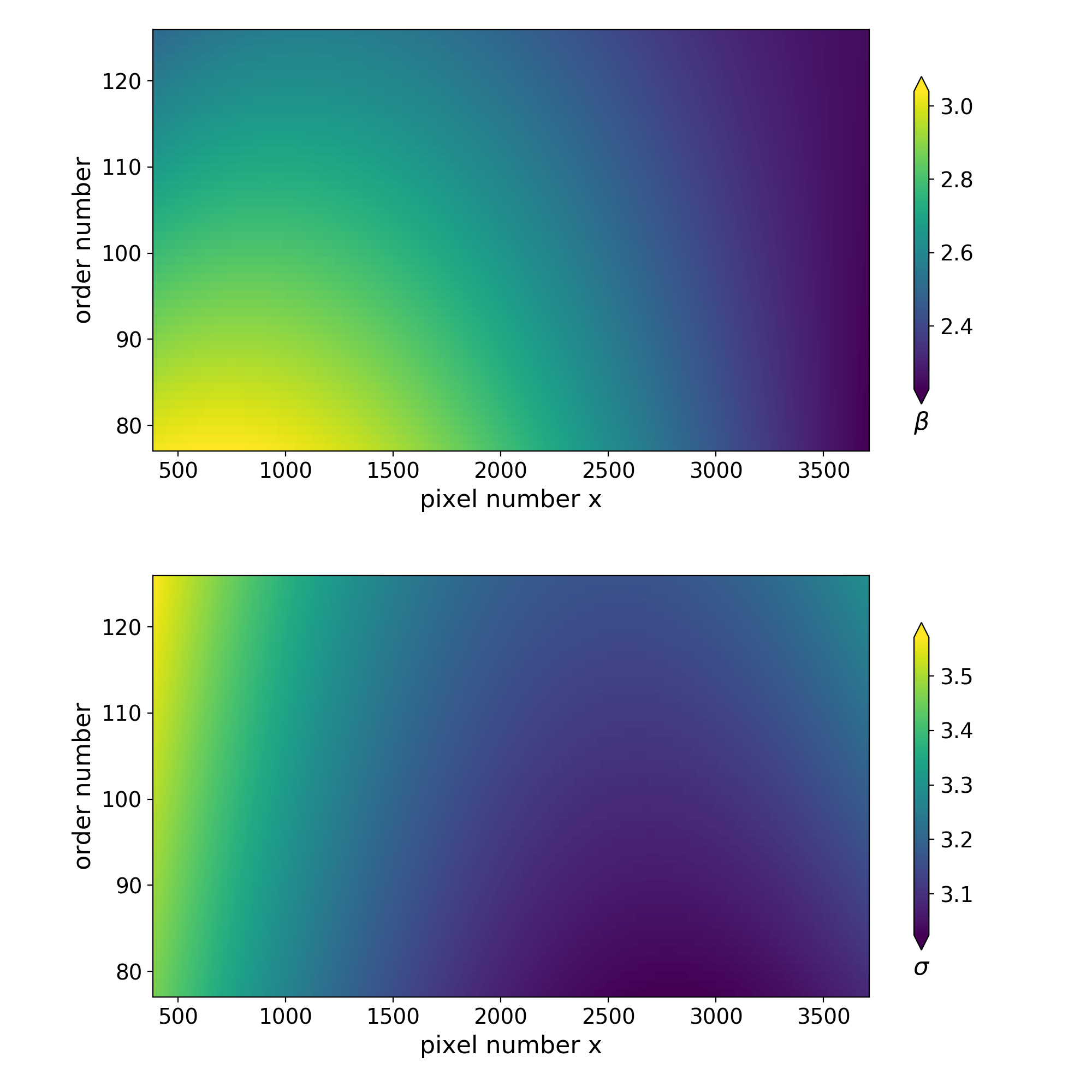}}
\caption{
Measured $\beta(o, x)$ (upper panel) and $\sigma(o, x)$ (lower panel) (the parameters of the backbone function as the functions of order $o$ and pixel position $x$) in the case of HRS.
The regions of $x<384$ and $x>3712$ are not plotted, since the lack of samples in these regions may cause unreliable results.
}
\label{fig:4.2}
\end{figure}

\begin{figure}
\centering
\resizebox{\hsize}{!}{\includegraphics{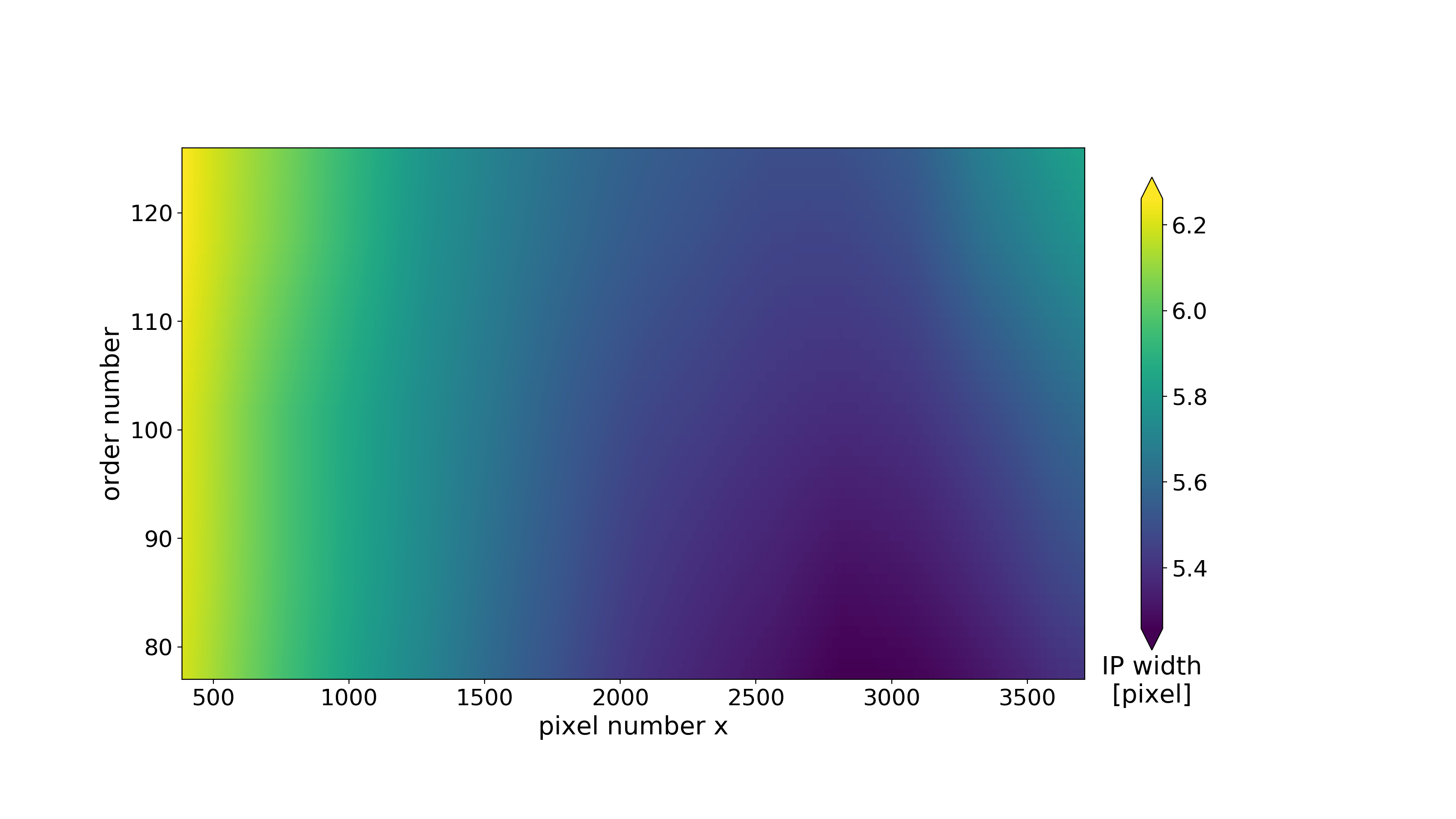}}
\caption{
The distribution of the width of the characterised IP in the case of HRS.
The regions of $x<384$ and $x>3712$ are not plotted, since the lack of samples in these regions may cause unreliable results.
}
\label{fig:4.3}
\end{figure}

We took four sequential exposures of ThAr lamp on HRS on September 22, 2017 and carried out the IP characterisation.
The exposure time was 10 s and the readout time was 200 s for each exposure.
When we implement the $\sigma(o, x)$ and $\beta(o, x)$ fitting and the residual function fitting, we combine the data points of all the four exposures.
For a single exposure, the average number of applicable ThAr lines in a block is about 15.
After combination, this average number increases to about 60.
 
The result of IP characterisation at nine positions on the CCD is shown in Fig.\ref{fig:4.1}.
These nine positions are marked in Fig.\ref{fig:3.2} with the red crosses. 
They correspond to the four corners, the centres of four sides, and the centre of the region of blocks, respectively.
We find that while the shape of IP varies slightly with the order $o$, it varies significantly with the pixel position $x$.
The spatial variation of IP along the cross-dispersion direction is induced by the optical aberrations.
Therefore, the result is consistent with the analysis in Section 2.1, where we argued the anamorphic tangential magnification produces much bigger spatial variation of IP than the optical aberrations.
The result also validates the way of our block division that the division along the principle dispersion direction needs to be denser than that along the cross-dispersion direction.

The result of the parameters of the backbone function -- $\beta(o, x)$ and $\sigma(o, x)$ -- is plotted in Fig.\ref{fig:4.2}.
$\beta(o, x)$ describes the distribution of the peakedness of IP.
A larger $\beta$ means a less peaked shape (i.e. closer to a flat-topped shape).
$\beta = 2$ is the peakedness of a Gaussian function.
The shape of a function with $\beta > 2$ is more flat-topped than the Gaussian function.
As shown in the upper panel of Fig.\ref{fig:4.2}, we see that $\beta(o, x) > 2$ everywhere over the entire range of CCD.
The lower panel of Fig.\ref{fig:4.2} shows the spatial variation of $\sigma(o, x)$, which roughly describes the width of IP.
But, more accurately, the width of IP correlates not only with $\sigma(o, x)$, but also with $\beta(o, x)$ and the residual function.
Therefore, we show the distribution of the width of the characterised IP in Fig.\ref{fig:4.3}.
We see that $\sigma(o, x)$ and the width of IP are both essentially monotonic declining towards the right side.
It is in accordance with the analysis of anamorphic tangential magnification in Section 2.1.
We also note that the slight increase near the upper right corner is possibly caused by the optical aberrations.
The average value of the width of the characterised IP is about 5.8 pixels, which is a little larger than the previously-measured value of 5.5 pixels by Gaussian function fitting. 

The result of the residual function is shown in Appendix A.
The three figures in Appendix A correspond to three rows of blocks, from bottom to top, respectively.
We see that the combination of residuals provides rich sampling rate in the fitting range, and the cubic spline function successfully captures the undulating relation of data.  
We also see that the spatial variation of the residual function from block to block is gradual, as we expect.
It demonstrates the good internal consistency of these combined residuals which appear as the same function. 
We note that the residual function of each block displays excellent convergence to zero inside [-7.5, 7.5] and rather flat shape around zero outside [-7.5, 7.5].
It demonstrates the influence of boundary condition is well controlled outside [-7.5, 7.5] by introducing the buffer spaces.
It is worth noting that the asymmetry of the residual function indicates the intrinsic asymmetry of the IP, given that the backbone function is symmetric.
It reveals that a symmetric function is unable to model the IP of HRS to a high accuracy.

\section{The accuracy of IP characterisation}

\begin{figure}
\centering
\resizebox{\hsize}{!}{\includegraphics{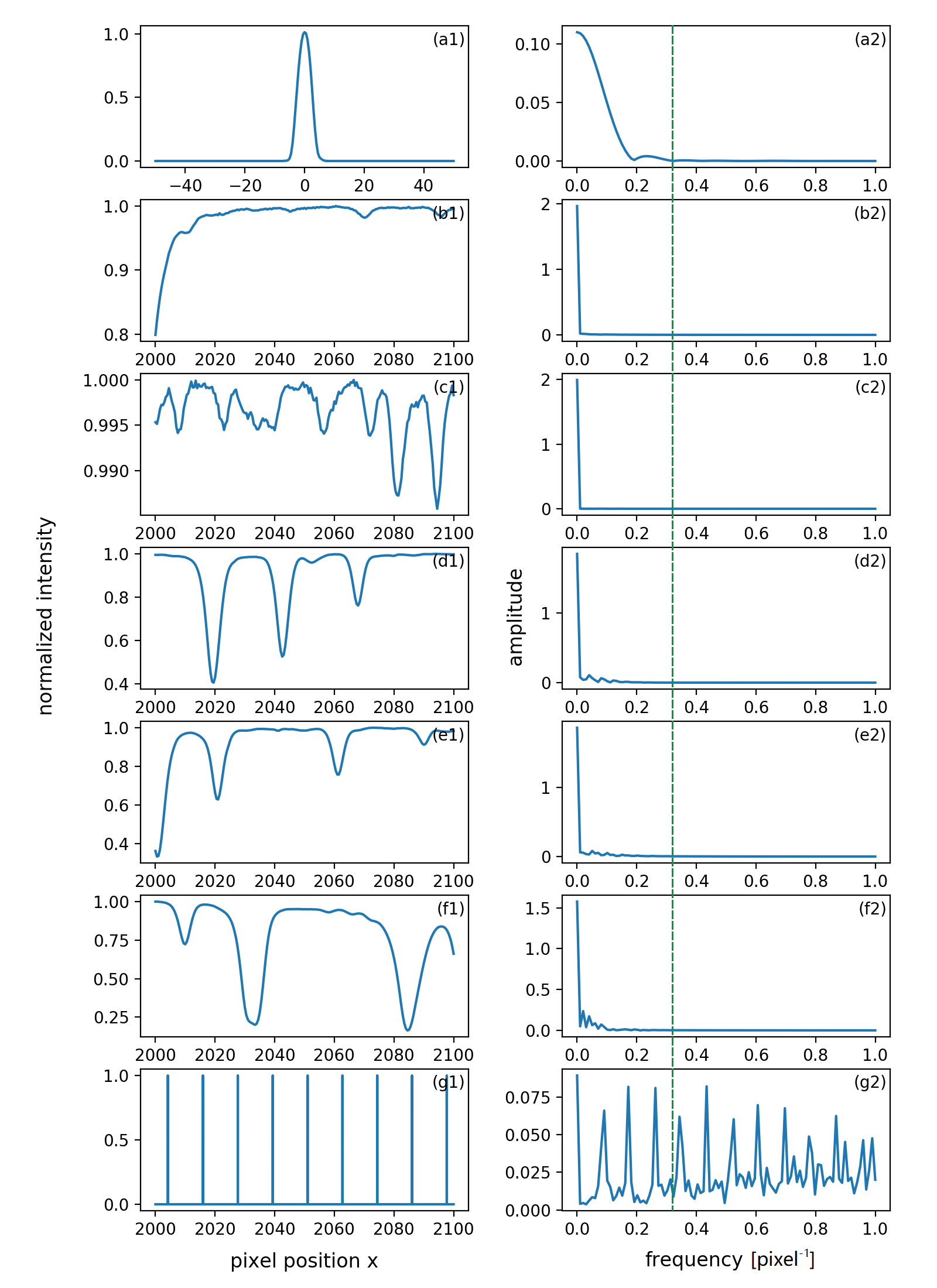}}
\caption{
{\bfseries a1)} The characterised IP in the centre ($x$=2048, $o$=100) of the CCD on HRS. 
{\bfseries b1-f1)} The very high resolution ($R$=300,000) template spectra of the sun \citep{kurucz1984} in five places on the axis ($x$=2000\textasciitilde2100, $o$=80, 90, 100, 110 and 120, respectively) of the CCD on HRS in the unit of pixel.
{\bfseries g1)} The template spectrum of the astro-comb in the centre ($x$=2000\textasciitilde2100, $o$=100) of the CCD on HRS in the unit of pixel. 
{\bfseries a2-g2)} Fourier transform of (a1)-(g1).
Note that the characterised IP and the template spectrum of the astro-comb are all converted to a resolution of 300,000. 
For (b1)-(g1), the unit of nm (wavelength) is converted to the unit of pixel based on the wavelength solution of HRS. 
The highest frequency of the characterised IP is marked with a vertical green line.
}
\label{fig:5.1}
\end{figure}

\begin{figure*}
\centering
\resizebox{\hsize}{!}{\includegraphics{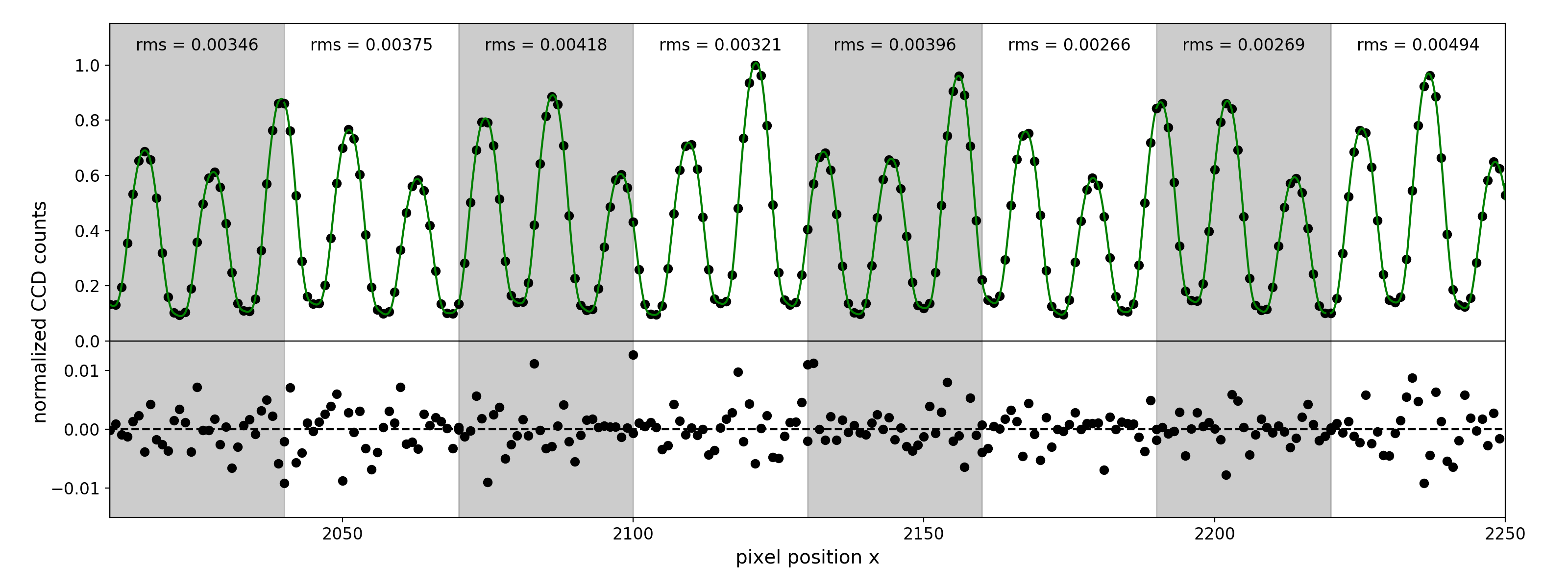}}
\caption{
Upper panel: Part of the normalised observed spectrum (black dots) and the reconstructed spectrum (green curve) of the astro-comb in order 100.
Lower panel: The corresponding residuals.
Different chunks are marked with different hue background.
The rms of each chunk is also marked.
}
\label{fig:5.2}
\end{figure*}

\begin{figure*}
\centering
\resizebox{\hsize}{!}{\includegraphics{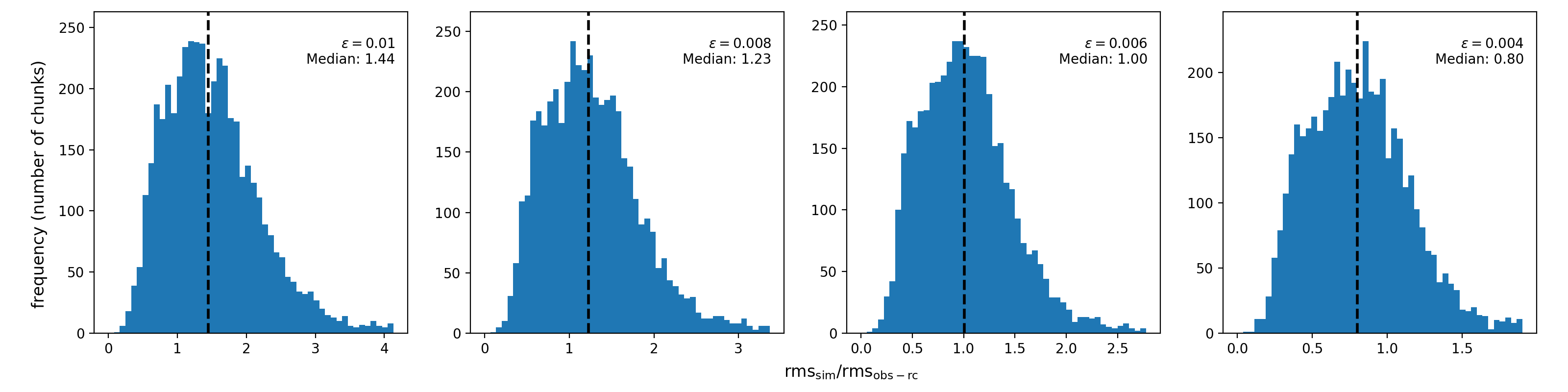}}
\caption{
Monte Carlo simulation result of $\textrm{rms}_\textrm{sim} / \textrm{rms}_\textrm{obs-rc}$ with respect to different $\epsilon$. 
From left to right, the histograms of $\textrm{rms}_\textrm{sim} / \textrm{rms}_\textrm{obs-rc}$ when $\epsilon = 0.01, 0.008, 0.006, 0.004$, respectively.
The median of the data of each histogram is marked by the dashed black line.
}
\label{fig:5.3}
\end{figure*}

\begin{figure*}
\centering
\resizebox{\hsize}{!}{\includegraphics{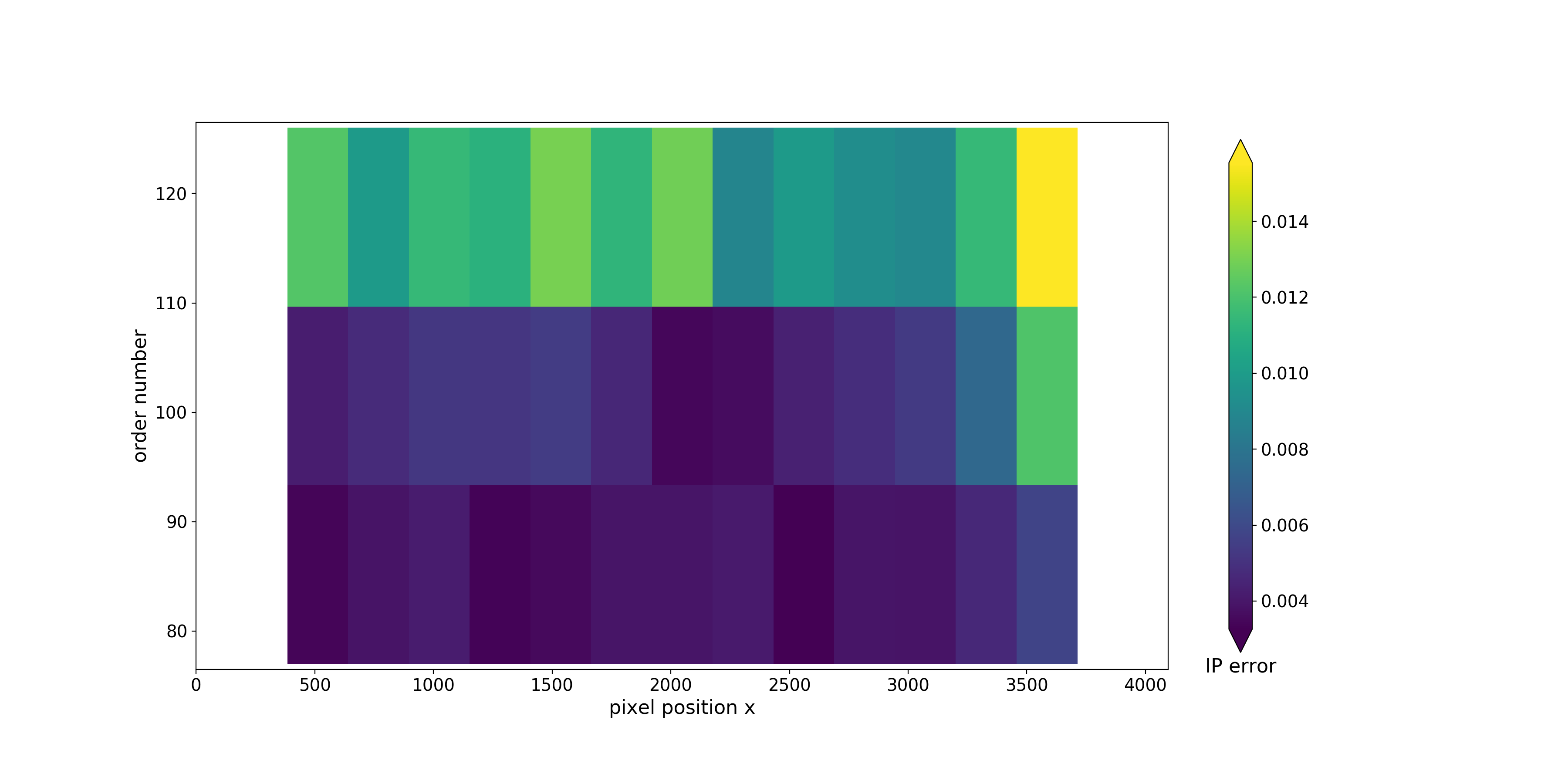}}
\caption{
The accuracy of IP characterisation with respect to different positions on the CCD.
The colour indicates the average error (accuracy) of IP characterisation in different blocks.
}
\label{fig:5.4}
\end{figure*}

\subsection{The method to determine the accuracy of IP characterisation}

Some information of the IP accuracy can be revealed by using the characterised IP to refit the applicable ThAr lines.
However, it cannot reveal the included systematic errors (e.g. the spurious broadening caused by the limited intrinsic line widths of ThAr lines) of the characterised IP. 
In principal, a better way to evaluate the IP accuracy needs to utilise another light source.
If we have a very high resolution template spectrum of a light source, then we can reconstruct the observed spectrum by convolving the template spectrum with the characterised IP. 
By analysing the difference between the reconstructed spectrum and the observed spectrum, we would get more complete information about the IP accuracy. 
(We call the spectral-reconstruction light source the {\bfseries RC source} below.)
Nevertheless, the effectiveness of this method is related to the characteristic of the template spectrum of RC source.
High effectiveness needs the RC source to have:
\begin{enumerate}
\item an accurate template spectrum;
\item broad wavelength coverage;
\item high complexity of the spectral structure. (i.e. Sufficient high frequency components. Note that the frequency here is not the optical frequency of light but about the Fourier transform of the spectrum. If the spectrum lacks high frequency components, the high frequency components in the IP would be smoothed out when convolved with the template spectrum.)
\end{enumerate}

Astro-comb is promising to be the most powerful wavelength calibrator aiding the search for earth-mass exoplanets around solar-mass stars in the habitable zone with the radial velocity method \citep{steinmetz2008, li2008, braje2008, wilken2012, phillips2012, ycas2012, probst2020}.
Astro-comb is able to provide a series of extremely narrow, dense, and regularly-spaced emission lines (also called comb teeth), with a broad wavelength coverage. 
The optical frequency of the comb teeth satisfies the relation of $f_{n} = f_{0} + n \times f_{\textrm{rep}}$, where $f_{0}$ denotes the offset frequency, $f_{\textrm{rep}}$ denotes the repetition rate (the mode spacing), and $n$ is an integer which denotes the mode number.
By locking to a standard radio-frequency reference, the frequency of comb teeth can reach an absolute accuracy of better than $10^{-12}$ (0.3 mm s$^{-1}$) \citep{doerr2012}, much finer than the RV amplitude of 9 cm s$^{-1}$ of the earth dragging the sun.
An astro-comb (Menlo Systems GmbH) was installed on HRS in 2016 \citep{ye2016, wu2016}. 
It has a repetition frequency of 25 GHz and an operation wavelength range of 470-720 nm, covering most of the visible region.
Using it to calibrate HRS, \cite{hao2018} demonstrated a 2-8 times (for different orders) higher wavelength solution accuracy than the ThAr lamp and a short-term repeatability of 0.1 m s$^{-1}$.
In principle, the extremely narrow, dense, and regularly-spaced comb teeth of the astro-comb are perfect for characterising the IP. 
However, HRS’s average resolving power of $4.3 \times 10^{4}$ cannot completely separate the 25-GHz-mode-spacing adjacent comb teeth, i.e. the comb teeth overlap with one another. 
So, the IP cannot be directly sampled using this astro-comb. 
In spite of it, due to the characteristic of the spectrum and the exact knowledge about the frequency of each comb tooth, the astro-comb is still the most favourable RC source on HRS.

For illustration, Fig.\ref{fig:5.1} shows the Fourier transform of the characterised IP, the template spectra of the sun and the astro-comb.
For the solar template spectra, we use the Kurucz solar flux atlas \citep{kurucz1984} which have a resolution of 300,000.
We casually select five segments of the solar template spectra which are on the central axis of the CCD.
The template spectrum of the astro-comb is a Delta-function sequence based on the formula of $f_{n} = f_{0} + n \times f_{\textrm{rep}}$ (simplified but enough here).
The characterised IP and the template spectrum of the astro-comb are both converted to a resolution of 300,000.
We can see that the highest frequency of the characterised IP is much higher than that of the five segments of solar template spectra.
It is probable that if we use a sun-like G-type star, or even any other astronomical object, as the RC source, lots of the information of the characterised IP would be lost after the convolution.
In contrast, the high frequency components of the spectrum of the astro-comb are very sufficient.
The high frequency components of the characterised IP can be preserved after the convolution.

Since the SNR for the observed spectrum is not infinite and the template spectrum is not perfect, the residuals between the observed spectrum and the reconstructed spectrum include random noise (photon noise, readout noise), errors in the template spectrum, errors of IP characterisation, etc. 
Simply using the residuals to represent the IP accuracy is not appropriate. 
Therefore, we propose the following procedure to strictly evaluate the accuracy of IP characterisation with the astro-comb as the RC source.

First, we divide the spectrum of the astro-comb into different chunks. 
In each chunk, the following equation links the observed spectrum and the original spectrum of RC source:
\begin{equation}
S_{\textrm{obs}}(x) = S(x) \convolution c\textrm{IP}_{\textrm{true}}(x) + n(x),
\label{eq:5.1}
\end{equation}
where $S_{\textrm{obs}}(x)$ is the observed spectrum, $S(x)$ is the original spectrum of the RC source, $\textrm{IP}_{\textrm{true}}(x)$ is the true IP of the spectrograph, $c$ is the normalisation constant (in our definition, the integral of IP is NOT equal to 1) and $n(x)$ is the random noise.

Here, we generate the complicated template spectrum with some free parameters to represent the original spectrum of the astro-comb.
For a chunk including $m$ comb teeth, the template spectrum is: 
\begin{equation}
S(x) = \sum_{i=1}^{m} a_{i}\delta(x-x_{\lambda}(\lambda_{i})) +  \sum_{i=0}^{n} b_{i}x^{i}.
\label{eq:5.2}
\end{equation}
$a_{i}$ (free parameter) and $\lambda_{i}$ are the intensity and the wavelength of the $i$th comb tooth, respectively.
$\lambda_{i}$ is exactly known, and can be obtained by the formula of $f_{n} = f_{0} + n \times f_{\textrm{rep}}$.
$x_{\lambda}(\lambda)$ is the inverse function of the wavelength solution of HRS. 
Delta function is used because the intrinsic line width of a comb tooth is extremely narrow (< 1 kHz, i.e. about $10^{-7}$ of the IP width) \citep{li2008}.
The polynomial term on the right side of Eq. \ref{eq:5.2} describes the continuum background in astro-comb's spectrum, the intensity of which is (1-3)\% of that of the comb teeth \citep{milakovic2017}. 
The coefficients $\{b_{i}\}_{i=0}^{n}$ are free parameters.
In our case, each chunk is 30-pixel wide and the cubic polynomial ($n=3$) is adopted. 
The spectral structure of the astro-comb is so regular that we ignore the errors in the template spectrum.

On the other hand, the true IP connects with the characterised IP as:
\begin{equation}
\textrm{IP}(x) =  \textrm{IP}_{\textrm{true}}(x) + \textrm{IP}_{\textrm{err}}(x),
\label{eq:5.3}
\end{equation}
where $\textrm{IP}(x)$ is the characterised IP and $\textrm{IP}_{\textrm{err}}(x)$ represents the errors of IP characterisation.
Putting Eq. \ref{eq:5.3} into Eq. \ref{eq:5.1}, we obtain:
\begin{equation}
\begin{split}
S_{\textrm{obs}}(x) =& \ S(x) \convolution c[\textrm{IP}(x) - \textrm{IP}_{\textrm{err}}(x)] + n(x) \\
                               =& \ S(x) \convolution c\textrm{IP}(x) - S(x) \convolution c\textrm{IP}_{\textrm{err}}(x) + n(x) \\
                               =& \ S_{\textrm{rc}}(x) - S(x) \convolution c\textrm{IP}_{\textrm{err}}(x) + n(x).
\label{eq:5.4}
\end{split}
\end{equation}
$S_{\textrm{rc}}(x)$ is the reconstructed spectrum based on the characterised IP.
It can be obtained by optimising $\{a_{i}\}_{i=0}^{m}$ and $\{b_{i}\}_{i=0}^{n}$ to minimise the weighted sum of the squares of $S_{\textrm{obs}}(x) - S_{\textrm{rc}}(x)$.
The weight for each pixel is equal to the reciprocal of the sum of the squares of the photon noise and the readout noise. 
So, the residuals between the observed spectrum and the reconstructed spectrum are contributed by the two terms on the right side:
\begin{equation}
S_{\textrm{obs}}(x) - S_{\textrm{rc}}(x) = - S(x) \convolution c\textrm{IP}_{\textrm{err}}(x) + n(x).
\label{eq:5.5}
\end{equation}
According to Eq. \ref{eq:5.5}, $\textrm{IP}_{\textrm{err}}(x)$ can be obtained with the deconvolution method.
And the errors (accuracy) of IP characterisation are thus found.

However, due to the calculational complication, we do not plan to use the deconvolution method to find $\textrm{IP}_{\textrm{err}}(x)$ here.
We solve it in a more straightforward way based on the Monte Carlo method.
When calculating $S(x) \convolution c\textrm{IP}_{\textrm{err}}(x)$ on the right side of Eq. \ref{eq:5.5}, we ignore the polynomial term of $S(x)$, which is a small quantity compared with $S(x)$. 
So, $S(x) \convolution c\textrm{IP}_{\textrm{err}}(x)$ is just the sum of (intensity-modulated) translated copies of $\textrm{IP}_{\textrm{err}}(x)$ at $S(x)$'s Delta functions' positions.
On the other hand, the characterised IP of HRS converges to zero inside [-7.5, 7.5].
It means $\textrm{IP}_{\textrm{err}}(x) = 0$ outside [-7.5, 7.5].
The interval of integration of the convolution thus shrinks to [-7.5, 7.5].
Since the comb-tooth spacing on HRS is 9-14 pixels, the interval of integration covers no more than two translated copies of $\textrm{IP}_{\textrm{err}}(x)$.
The convolution can be regarded as a successive process of sampling $\textrm{IP}_{\textrm{err}}(x)$ (once or twice) and summing up the samples.
Due to the ergodicity of the pixel phase of comb teeth, the sampling process can be assumed to be statistically random. 
Therefore, based on the Monte Carlo method, we can simulate $S(x) \convolution c\textrm{IP}_{\textrm{err}}(x)$ with random variables:
\begin{displaymath}
\begin{split}
& S(x) \convolution c\textrm{IP}_{\textrm{err}}(x) \rightarrow a_{j}c\xi_{1}, \\
& (\textrm{if the } j \textrm{th comb tooth is in } [x-7.5, x+7.5]) \\
& S(x) \convolution c\textrm{IP}_{\textrm{err}}(x) \rightarrow a_{j}c\xi_{1} + a_{j+1}c\xi_{2}. \\
& (\textrm{if the } j \textrm{th and (} j \textrm{+1)th comb teeth are both in } [x-7.5, x+7.5]) 
\label{eq:5.6}
\end{split}
\end{displaymath}
$\xi_{1}$ and $\xi_{2}$ are independent random variables, and they are both distributed normally with mean 0 and standard deviation $\epsilon$: $\xi_{1} \thicksim N(0, \epsilon^{2})$ and $\xi_{2} \thicksim N(0, \epsilon^{2})$.
$\epsilon$ is the standard deviation of $\textrm{IP}_{\textrm{err}}(x)$.
$a_{j}$ and $a_{j+1}$ are the optimised parameters in the reconstructed spectrum.

The random noise in Eq. \ref{eq:5.5} is also converted to a random variable.
It is distributed normally with mean 0 and standard deviation $\sigma$: $n(x) \thicksim N(0, \sigma^{2})$, where $\sigma^{2}$ is the sum of the square of the photon noise and the readout noise: $\sigma^{2}=S_{\textrm{obs}}(x) + \textrm{RN}^{2}$.

So far, we define the simulated residual between the observed spectrum and the reconstructed spectrum at $x$ as:
\begin{displaymath}
\begin{split}
& \textrm{res}_\textrm{sim}(x) = -a_{j}c\xi_{1} + n(x), \\
& (\textrm{if the } j \textrm{th comb tooth is in } [x-7.5, x+7.5]) \\
& \textrm{res}_\textrm{sim}(x) = -a_{j}c\xi_{1} - a_{j+1}c\xi_{2} + n(x), \\
& (\textrm{if the } j \textrm{th and (} j \textrm{+1)th comb teeth are both in } [x-7.5, x+7.5]) 
\label{eq:5.7}
\end{split}
\end{displaymath}
and the actual residual between the observed spectrum and the reconstructed spectrum at $x$ as:
\begin{displaymath}
\textrm{res}_\textrm{obs-rc} = S_{\textrm{obs}}(x) - S_{\textrm{rc}}(x).
\label{eq:5.8}
\end{displaymath}
For each chunk, the rms values of $\textrm{res}_\textrm{sim}$ and $\textrm{res}_\textrm{obs-rc}$ are, respectively:
\begin{displaymath}
\begin{split}
& \textrm{rms}_\textrm{sim} = \frac{1}{30} \sum\limits_{x} \textrm{res}_\textrm{sim}(x), \\
& \textrm{rms}_\textrm{obs-rc} = \frac{1}{30} \sum\limits_{x} \textrm{res}_\textrm{obs-rc}(x).
\label{eq:5.9}
\end{split}
\end{displaymath}
We implement the simulation (generate random inputs) with the Python programming language, and collect the ratios of $\textrm{rms}_\textrm{sim}$ to $\textrm{rms}_\textrm{obs-rc}$ of all chunks.
If $\epsilon$, the standard deviation of $\textrm{IP}_{\textrm{err}}(x)$, is correctly set, the median of them should satisfy:
\begin{equation}
\textrm{MEDIAN} (\frac{\textrm{rms}_\textrm{sim}}{\textrm{rms}_\textrm{obs-rc}}) = 1.
\label{eq:5.10}
\end{equation}
This $\epsilon$ is just the average error (accuracy) of IP characterisation on HRS.

\subsection{The result of the accuracy of IP characterisation}

We took an exposure of the astro-comb on HRS immediately after the four exposures of ThAr lamp on September 22, 2017.
Following the procedure in Section 5.1, We use this exposure to evaluate the accuracy of the characterised IP in Section 4.

The spectral reconstruction is an important intermediate step of the procedure in Section 5.1.
Fig.\ref{fig:5.2} gives an example of the result of spectral reconstruction based on the characterised IP.
The reconstructed spectrum in this part is very accurate, as the rms of each chunk is all well below 0.005 of the maximum of this part in order 100.

Fig.\ref{fig:5.3} shows the simulation result of $\textrm{rms}_\textrm{sim} / \textrm{rms}_\textrm{obs-rc}$ with respect to different $\epsilon$. 
We plot the histograms of $\textrm{rms}_\textrm{sim} / \textrm{rms}_\textrm{obs-rc}$ for $\epsilon = 0.01, 0.008, 0.006, 0.004$, and we find the medians of them are 1.44, 1.23, 1.00, 0.80, respectively.
It means the average error (accuracy) of the characterised IP is 0.006 of the peak value of the backbone function.
The result demonstrates that accurate IP characterisation has been achieved with the BR model and the ThAr lines on HRS.

The accuracy of IP characterisation is not homogeneous everywhere, but varies across the CCD.
Fig.\ref{fig:5.4} shows the relation between the accuracy of IP characterisation and the position on the CCD.
The colour indicates the average error (accuracy) of IP characterisation in different blocks.
Obviously, the quantity and the distribution of ThAr lines will influence the IP-characterisation accuracy.
For example, the comparatively bad accuracy in the top right corner of the region of blocks could be caused by the non-uniform distribution of ThAr lines.
We can see from Fig.\ref{fig:3.2} that the applicable ThAr lines in this block gather together in the lower right corner.
On the other hand, we also see that better IP-characterisation accuracy is obtained in lower orders while worse in higher orders.
However, the majority of those chunks have similar quantity and distribution of ThAr lines.
We exclude the anomaly of the continuum background in astro-comb's spectrum in the higher orders as the main reason for this trend, because no improvement is accomplished after we increase the degree of the polynomial in Eq. \ref{eq:5.2} for a better continuum fitting.
This leads us to speculate that the reason is more likely the worse accuracy of wavelength solution for the higher orders \citep{hao2018}.
The true nature of such a phenomenon needs to be scrutinised but it is beyond the scope of the present work. 
We leave a detailed investigation on this issue to the future study.
Nevertheless, it is more likely that the worse IP-characterisation accuracy in the higher orders is caused by the imperfect algorithm of the procedure in Section 5.1.
The actual accuracy would be at the same level of that in the lower orders.
If we average the accuracy of IP characterisation in the lower two rows of blocks, an IP-characterisation accuracy of 0.0047 can be obtained.
This level approaches the theoretical limit under our selection criterion for the ThAr lines.
It further demonstrates the excellence of the BR model.

\section{Discussion and Conclusions}

In this paper, we investigate the IP characterisation for the fibre-fed echelle spectrograph using the emission lines of the ThAr lamp as the IP samples.
We clarify the selection criterion for the ThAr lines, and find the anamorphic tangential magnification to be the main factor causing the spatial variation of IP on the CCD.
Based on this, we propose the backbone-residual (BR) model to characterise the IP. 
A bell-shaped function acts as the backbone function describing the main component and the spatial variation of IP. 
The residual function, which is expressed as the cubic spline function, accounts for the difference between the bell-shaped function and the actual IP. 
We verified this method on HRS at the Chinese Xinglong 2.16-m Telescope, and obtained the characterised IP.
To evaluate the BR model, we propose a procedure to calculate the IP-characterisation accuracy by using the astro-comb, based on the spectral reconstruction and Monte-Carlo simulation. 
And we conclude that the average accuracy of IP characterisation on HRS is 0.006 of the peak value of the backbone function.
This result demonstrates that the BR model is an excellent choice in accurate IP characterisation.

We employ the normalised super Gaussian function as the backbone function based on the actual shape of the IP of HRS.
In principle, finding a function form to better capture the shape and the spatial variation of IP is generally preferable.
But practically, the choice of the backbone function would not influence the final IP-characterisation accuracy much due to the residual function involved. 
Instead, we need to prevent overfitting by avoiding the employment of an excessively complicated function as the backbone function.
Owing to the small amount of applicable ThAr lines close to the edges of the CCD (the result of the modulation of the blaze function), we do not characterise the IP in this region. 
To realise accurate IP characterisation in this region, one could co-add multiple exposures of ThAr lamp to increase the SNR of ThAr lines.
However, it should be noted that the spectrograph drift would lead to the deformation of IP samples in co-added exposures.
Our procedure to evaluate the IP-characterisation accuracy is not only limited to the fibre-fed echelle spectrographs with astro-combs, but can be easily generalised to other RC sources featuring emission lines.
For the RC sources featuring absorption lines, our framework of analysis is still applicable, but the IP-characterisation accuracy will need to be solved by the deconvolution method instead of the Monte Carlo simulation.

In recent years, newly-developed next generation calibration sources are equipped or tested on more and more fibre-fed echelle spectrographs all over the world. 
The next generation calibration sources are also powerful tools to characterise the IP of spectrographs. 
For example, the astro-combs and Fabry-P\'erot (F-P) etalons are promising to describe the IP more accurately than ThAr lamps with the help of their regularly-spaced and densely-distributed comb teeth. 
This makes possible the accurate IP characterisation on the next generation exoplanet hunters (e.g. ESPRESSO, G-CLEF, EXPRES), aiding the advanced astronomical researches such as the measurements of extremely precise radial velocity and potential cosmological variability of fundamental constants. 
The BR model could play a role in this area.
However, there are still some limits that would complexify the procedure to characterise the IP.
For example, for F-P etalons, due to the limited finesse, the intrinsic line shape of comb teeth would significantly deform the true IP of the spectrograph. 
Further studies are necessary in the future.

\begin{acknowledgements}
We acknowledge the support of the National Natural Science Foundation of China (NSFC) (Grant Nos. 11803060, 11673046, 11727806, 11773044, 11773047, U1831209, 11703052), the International Partnership Program of Chinese Academy of Sciences (Grant No. 114A32KYSB20160049) and the National Key R\&D Program of China (Grant No. 2016YFA0400800).
We acknowledge the support of the staff of the Xinglong 2.16-m telescope. 
This work was partially supported by the Open Project Program of the CAS Key Laboratory of Optical Astronomy, National Astronomical Observatories.
We acknowledge Tilo Steinmetz, Yuanjie Wu and other staff members of Menlo System GmbH for the maintenance of the astro-comb.
\end{acknowledgements}

\bibliographystyle{aa}
\bibliography{ref}

\begin{thebibliography}{31}
\expandafter\ifx\csname natexlab\endcsname\relax\def\natexlab#1{#1}\fi

\bibitem[{{Anderson} \& {King}(2000)}]{anderson2000}
{Anderson}, J. \& {King}, I.~R. 2000, \pasp, 112, 1360

\bibitem[{{Bland-Hawthorn} {et~al.}(2017){Bland-Hawthorn}, {Kos}, {Betters},
  {De Silva}, {O'Byrne}, {Patterson}, \& {Leon-Saval}}]{bh2017}
{Bland-Hawthorn}, J., {Kos}, J., {Betters}, C.~H., {et~al.} 2017, Optics
  Express, 25, 15614

\bibitem[{{Bolton} \& {Schlegel}(2010)}]{bolton2010}
{Bolton}, A.~S. \& {Schlegel}, D.~J. 2010, \pasp, 122, 248

\bibitem[{{Braje} {et~al.}(2008){Braje}, {Kirchner}, {Osterman}, {Fortier}, \&
  {Diddams}}]{braje2008}
{Braje}, D.~A., {Kirchner}, M.~S., {Osterman}, S., {Fortier}, T., \& {Diddams},
  S.~A. 2008, European Physical Journal D, 48, 57

\bibitem[{{Buchhave}(2010)}]{buchhave2010}
{Buchhave}, L.~A. 2010, PhD thesis, University of Copenhagen

\bibitem[{{Cornachione} {et~al.}(2019){Cornachione}, {Bolton}, {Eastman},
  {Wilson}, {Wang}, {Johnson}, {Sliski}, {McCrady}, {Wright}, {Plavchan},
  {Johnson}, {Horner}, \& {Wittenmyer}}]{cornachione2019}
{Cornachione}, M.~A., {Bolton}, A.~S., {Eastman}, J.~D., {et~al.} 2019, \pasp,
  131, 124503

\bibitem[{{Dekker} {et~al.}(2000){Dekker}, {D'Odorico}, {Kaufer}, {Delabre}, \&
  {Kotzlowski}}]{dekker2000}
{Dekker}, H., {D'Odorico}, S., {Kaufer}, A., {Delabre}, B., \& {Kotzlowski}, H.
  2000, in Society of Photo-Optical Instrumentation Engineers (SPIE) Conference
  Series, Vol. 4008, \procspie, ed. M.~{Iye} \& A.~F. {Moorwood}, 534--545

\bibitem[{{Doerr} {et~al.}(2012){Doerr}, {Steinmetz}, {Holzwarth},
  {Kentischer}, \& {Schmidt}}]{doerr2012}
{Doerr}, H.~P., {Steinmetz}, T., {Holzwarth}, R., {Kentischer}, T., \&
  {Schmidt}, W. 2012, \solphys, 280, 663

\bibitem[{{Fan} {et~al.}(2016){Fan}, {Wang}, {Jiang}, {Wu}, {Li}, {Huang},
  {Xu}, {Hu}, {Zhu}, {Wang}, {Komossa}, \& {Zhang}}]{fan2016}
{Fan}, Z., {Wang}, H., {Jiang}, X., {et~al.} 2016, \pasp, 128, 115005

\bibitem[{{Guangwei} {et~al.}(2015){Guangwei}, {Haotong}, \&
  {Zhongrui}}]{li2015}
{Guangwei}, L., {Haotong}, Z., \& {Zhongrui}, B. 2015, \pasp, 127, 552

\bibitem[{{Hao} {et~al.}(2018){Hao}, {Ye}, {Han}, {Wu}, {Zhai}, \&
  {Xiao}}]{hao2018}
{Hao}, Z., {Ye}, H., {Han}, J., {et~al.} 2018, \pasp, 130, 125001

\bibitem[{{Kos} {et~al.}(2018){Kos}, {Bland-Hawthorn}, {Betters}, {Leon-Saval},
  {Asplund}, {Buder}, {Casey}, {D'Orazi}, {de Silva}, {Freeman}, {Lewis},
  {Lin}, {Martell}, {Schlesinger}, {Sharma}, {Simpson}, {Zucker}, {Zwitter},
  {Hayden}, {Horner}, {Nataf}, \& {Ting}}]{kos2018}
{Kos}, J., {Bland-Hawthorn}, J., {Betters}, C.~H., {et~al.} 2018, \mnras, 480,
  5475

\bibitem[{{Kurucz} {et~al.}(1984){Kurucz}, {Furenlid}, {Brault}, \&
  {Testerman}}]{kurucz1984}
{Kurucz}, R.~L., {Furenlid}, I., {Brault}, J., \& {Testerman}, L. 1984, {Solar
  flux atlas from 296 to 1300 nm}

\bibitem[{{Li} {et~al.}(2008){Li}, {Benedick}, {Fendel}, {Glenday},
  {K{\"a}rtner}, {Phillips}, {Sasselov}, {Szentgyorgyi}, \&
  {Walsworth}}]{li2008}
{Li}, C.-H., {Benedick}, A.~J., {Fendel}, P., {et~al.} 2008, \nat, 452, 610

\bibitem[{{Martin} {et~al.}(2005){Martin}, {Ibata}, {Conn}, {Irwin}, \&
  {Lewis}}]{martin2005}
{Martin}, N.~F., {Ibata}, R.~A., {Conn}, B.~C., {Irwin}, M.~J., \& {Lewis},
  G.~F. 2005, \pasa, 22, 236

\bibitem[{{Mayor} {et~al.}(2003){Mayor}, {Pepe}, {Queloz}, {Bouchy},
  {Rupprecht}, {Lo Curto}, {Avila}, {Benz}, {Bertaux}, {Bonfils}, {Dall},
  {Dekker}, {Delabre}, {Eckert}, {Fleury}, {Gilliotte}, {Gojak}, {Guzman},
  {Kohler}, {Lizon}, {Longinotti}, {Lovis}, {Megevand}, {Pasquini}, {Reyes},
  {Sivan}, {Sosnowska}, {Soto}, {Udry}, {van Kesteren}, {Weber}, \&
  {Weilenmann}}]{mayor2003}
{Mayor}, M., {Pepe}, F., {Queloz}, D., {et~al.} 2003, The Messenger, 114, 20

\bibitem[{{Milakovic} {et~al.}(2017){Milakovic}, {Pasquini}, {Lo Curto},
  {Avila}, {Manescau}, {Canto Martins}, {Le{\~a}o}, {De Medeiros}, {Esposito},
  {Gonz{\'a}lez Hern{\'a}ndez}, {Rebolo}, {Probst}, {Steinmetz}, {H{\"a}nsch},
  {Udem}, \& {Holzwarth}}]{milakovic2017}
{Milakovic}, D., {Pasquini}, L., {Lo Curto}, G., {et~al.} 2017, in ESO
  Calibration Workshop: The Second Generation VLT Instruments and Friends, 30

\bibitem[{{Palmer} \& {Loewen}(2005)}]{christopher2005}
{Palmer}, C. \& {Loewen}, E. 2005, DIFFRACTION GRATING HANDBOOK

\bibitem[{{Phillips} {et~al.}(2012){Phillips}, {Glenday}, {Li}, {Cramer},
  {Furesz}, {Chang}, {Benedick}, {Chen}, {K{\"a}rtner}, {Korzennik},
  {Sasselov}, {Szentgyorgyi}, \& {Walsworth}}]{phillips2012}
{Phillips}, D.~F., {Glenday}, A.~G., {Li}, C.-H., {et~al.} 2012, Optics
  Express, 20, 13711

\bibitem[{{Probst} {et~al.}(2020){Probst}, {Milakovi{\'c}},
  {Toledo-Padr{\'o}n}, {Lo Curto}, {Avila}, {Brucalassi}, {Canto Martins}, {de
  Castro Le{\~a}o}, {Esposito}, {Gonz{\'a}lez Hern{\'a}ndez}, {Grupp},
  {H{\"a}nsch}, {Kellermann}, {Kerber}, {Mandel}, {Manescau}, {Pozna},
  {Rebolo}, {de Medeiros}, {Steinmetz}, {Su{\'a}rez Mascare{\~n}o}, {Udem},
  {Urrutia}, {Wu}, {Pasquini}, \& {Holzwarth}}]{probst2020}
{Probst}, R.~A., {Milakovi{\'c}}, D., {Toledo-Padr{\'o}n}, B., {et~al.} 2020,
  Nature Astronomy [\eprint[arXiv]{2002.08868}]

\bibitem[{{Redman} {et~al.}(2014){Redman}, {Nave}, \&
  {Sansonetti}}]{redman2014}
{Redman}, S.~L., {Nave}, G., \& {Sansonetti}, C.~J. 2014, \apjs, 211, 4

\bibitem[{{Schroeder}(1970)}]{schroeder1970}
{Schroeder}, D.~J. 1970, \pasp, 82, 1253

\bibitem[{{Schroeder}(1987)}]{schroeder1987}
{Schroeder}, D.~J. 1987, {Astronomical optics}

\bibitem[{{Steinmetz} {et~al.}(2008){Steinmetz}, {Wilken}, {Araujo-Hauck},
  {Holzwarth}, {H{\"a}nsch}, {Pasquini}, {Manescau}, {D'Odorico}, {Murphy},
  {Kentischer}, {Schmidt}, \& {Udem}}]{steinmetz2008}
{Steinmetz}, T., {Wilken}, T., {Araujo-Hauck}, C., {et~al.} 2008, Science, 321,
  1335

\bibitem[{{Valenti} {et~al.}(1995){Valenti}, {Butler}, \&
  {Marcy}}]{valenti1995}
{Valenti}, J.~A., {Butler}, R.~P., \& {Marcy}, G.~W. 1995, \pasp, 107, 966

\bibitem[{{Wang}(2016)}]{wang2016}
{Wang}, X. 2016, PhD thesis, Pennsylvania State University

\bibitem[{{Wilken} {et~al.}(2012){Wilken}, {Curto}, {Probst}, {Steinmetz},
  {Manescau}, {Pasquini}, {Gonz{\'a}lez Hern{\'a}ndez}, {Rebolo}, {H{\"a}nsch},
  {Udem}, \& {Holzwarth}}]{wilken2012}
{Wilken}, T., {Curto}, G.~L., {Probst}, R.~A., {et~al.} 2012, \nat, 485, 611

\bibitem[{{Wu} {et~al.}(2016){Wu}, {Ye}, {Han}, {Zou}, {Fu}, \&
  {Xiao}}]{wu2016}
{Wu}, Y., {Ye}, H., {Han}, J., {et~al.} 2016, Acta Optica Sinica, 36, 0614001

\bibitem[{{Ycas} {et~al.}(2012){Ycas}, {Quinlan}, {Diddams}, {Osterman},
  {Mahadevan}, {Redman}, {Terrien}, {Ramsey}, {Bender}, {Botzer}, \&
  {Sigurdsson}}]{ycas2012}
{Ycas}, G.~G., {Quinlan}, F., {Diddams}, S.~A., {et~al.} 2012, Optics Express,
  20, 6631

\bibitem[{{Ye} {et~al.}(2016){Ye}, {Han}, {Wu}, \& {Xiao}}]{ye2016}
{Ye}, H., {Han}, J., {Wu}, Y., \& {Xiao}, D. 2016, in Society of Photo-Optical
  Instrumentation Engineers (SPIE) Conference Series, Vol. 9908, \procspie,
  99087E

\bibitem[{{Zhao} {et~al.}(2014){Zhao}, {Zhao}, {Lo Curto}, {Wang}, {Liu},
  {Wang}, \& {Wang}}]{zhao2014}
{Zhao}, F., {Zhao}, G., {Lo Curto}, G., {et~al.} 2014, Research in Astronomy
  and Astrophysics, 14, 1037

\end{thebibliography}

\begin{appendix}
\section{The result of the residual function}

Figures in this appendix show the result of the best-fit residual function of each block. 
The best-fit residual function of each block is shown by the green curve.
The pixel position $x$ of the centre of each block is marked on the top of each subfigure. 
The data points in each subfigure are the combined (centred and normalised) residuals of applicable ThAr lines in each block, except for those in the buffer spaces (outside the interval of $x' \in [-7.5, 7.5]$) which are the artificially-arranged zeros.
The light blue ones are the data used to fit, and the red ones are the discarded points by 3$\sigma$-clipping. 
The error bars are based on the normalised uncertainties induced by the photon noise and the readout noise. 
The dashed lines in each subfigure mark the positions of $x'=\pm7.5$.
The 21 interior knots of the cubic spline function are -8.75, -7.5, -6.25, -5.25, -4.5, -3.75, -3.0, -2.25, -1.5, -0.75, 0.0, 0.75, 1.5, 2.25, 3.0, 3.75, 4.5, 5.25, 6.25, 7.5 and 8.75.

\begin{figure*}
\centering
\resizebox{\hsize}{!}{\includegraphics{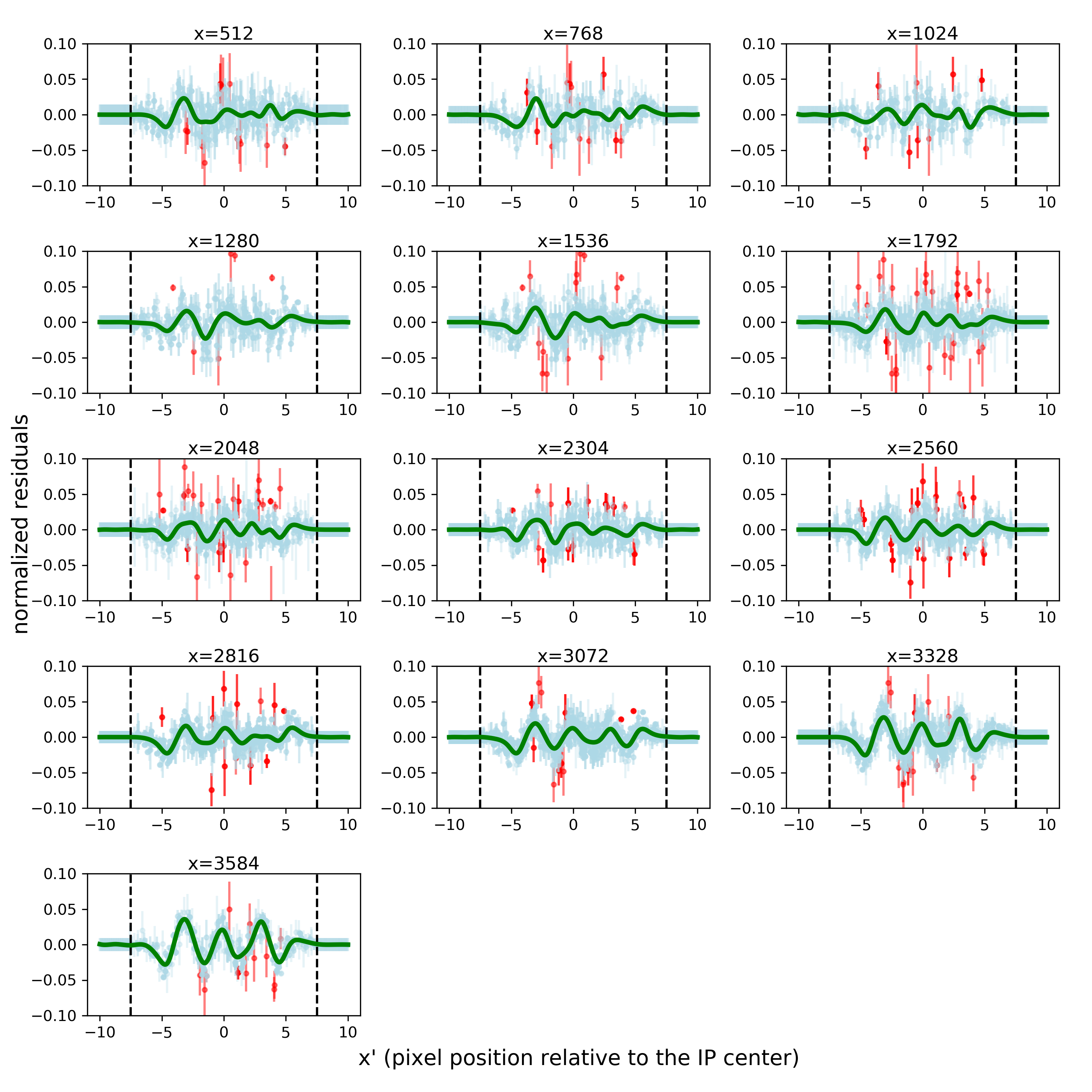}}
\caption{
The best-fit residual function of each block in the lower row in Fig.\ref{fig:3.2}.
}
\label{fig:A1}
\end{figure*}

\begin{figure*}
\centering
\resizebox{\hsize}{!}{\includegraphics{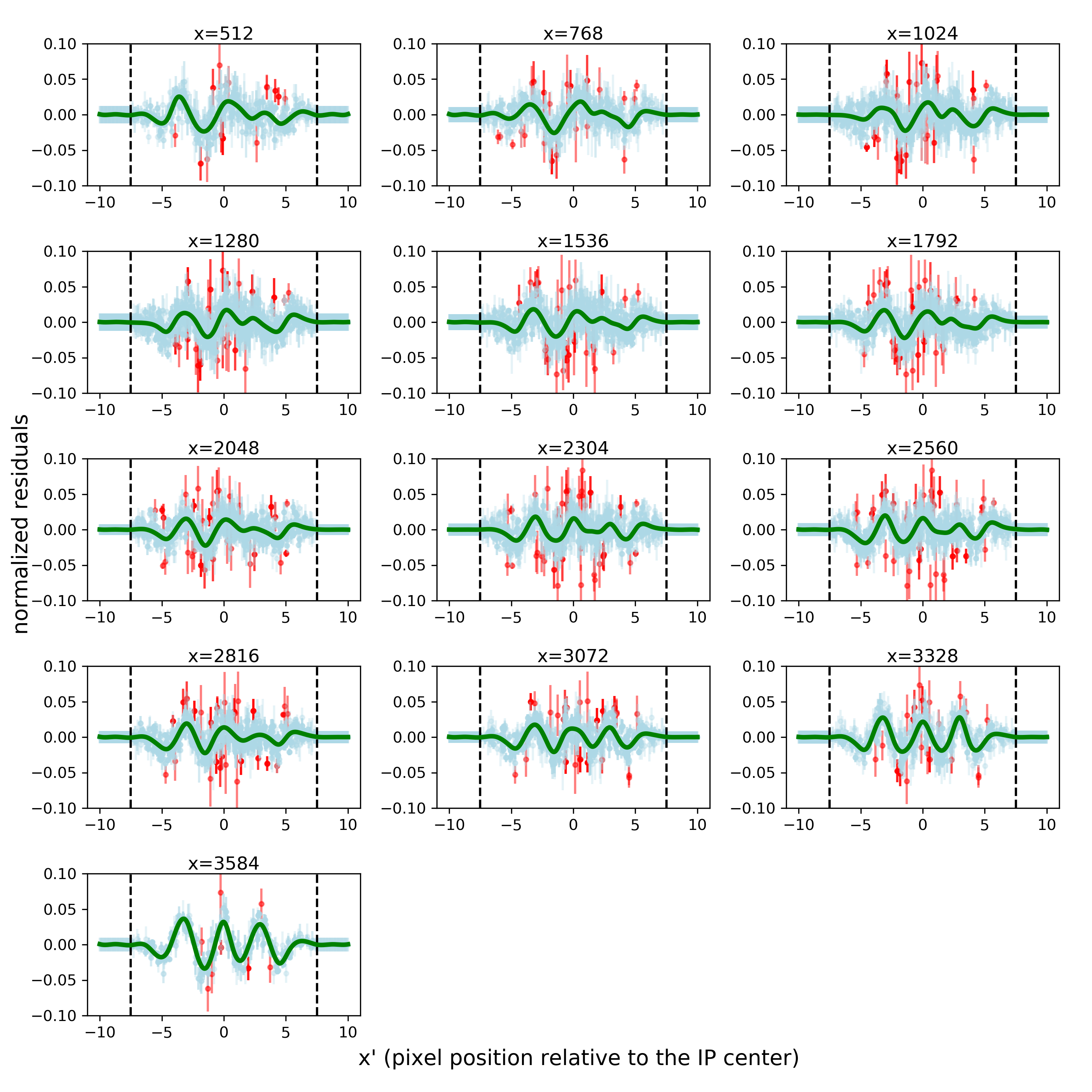}}
\caption{
The best-fit residual function of each block in the middle row in Fig.\ref{fig:3.2}.
}
\label{fig:A2}
\end{figure*}

\begin{figure*}
\centering
\resizebox{\hsize}{!}{\includegraphics{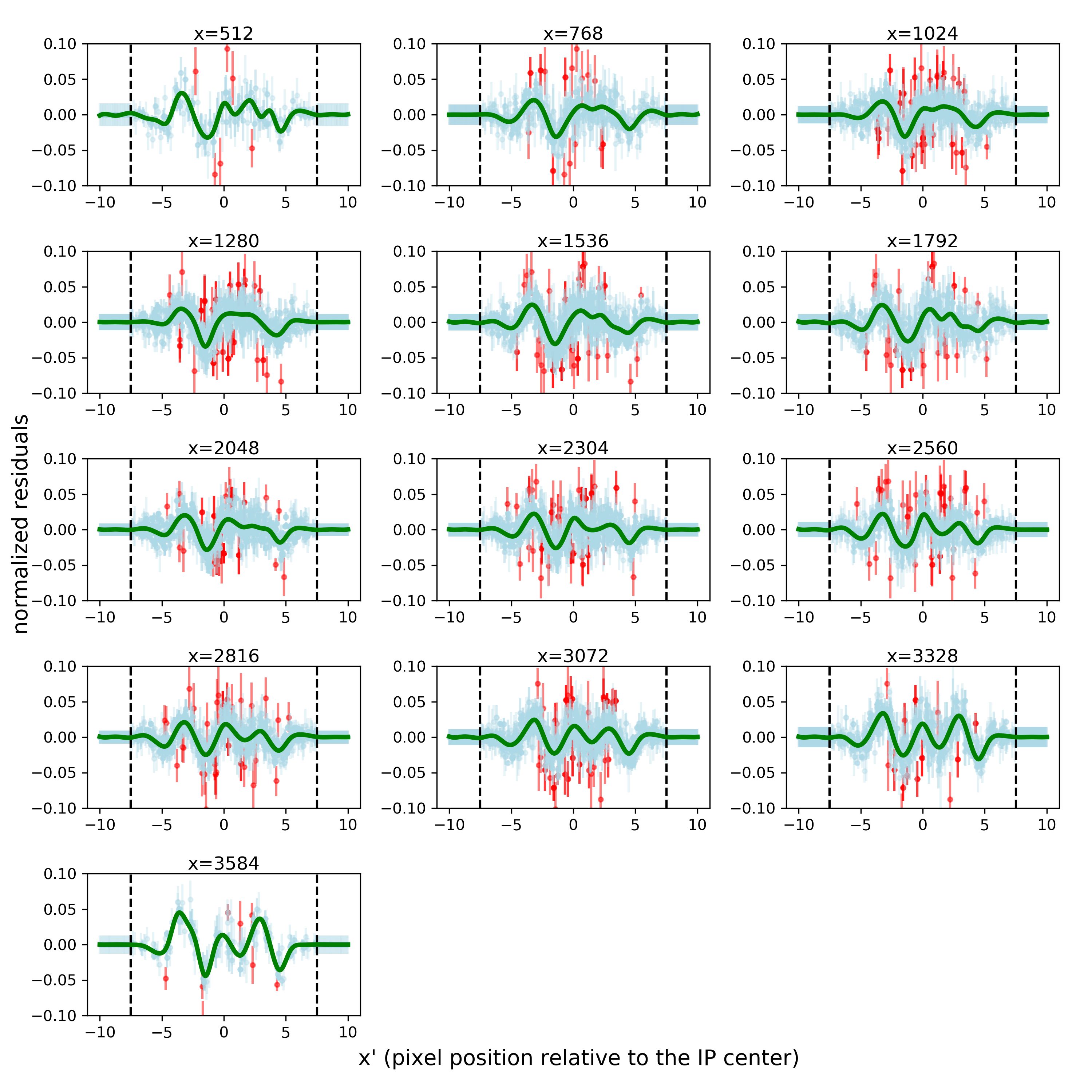}}
\caption{
The best-fit residual function of each block in the upper row in Fig.\ref{fig:3.2}.
}
\label{fig:A3}
\end{figure*}

\end{appendix}

\end{document}